\documentclass[final]{article}

\usepackage{arxiv}

\usepackage[utf8]{inputenc} 
\usepackage[T1]{fontenc}    
\usepackage{hyperref}       
\usepackage{url}            
\usepackage{booktabs}       
\usepackage{amsfonts}       
\usepackage{nicefrac}       
\usepackage{microtype}      
\usepackage{lipsum}
\usepackage{graphicx}

\usepackage{xcolor}
\usepackage{tikz}
\usepackage{amsmath}
\usepackage{amssymb}

\usepackage{lineno}
\usepackage{algorithm2e}
\usepackage{textcomp} 
\usepackage{subcaption}
\usepackage{adjustbox}

\modulolinenumbers[5]
\usetikzlibrary{calc}
\usetikzlibrary{intersections}
\usetikzlibrary{through}

\usepackage{nomencl}
\usepackage{etoolbox}

\makenomenclature

\usepackage{etoolbox}
\renewcommand\nomgroup[1]{%
  \item[\bfseries
  \ifstrequal{#1}{A}{Abbreviations}{}%
  \ifstrequal{#1}{F}{Feature Extraction Symbols}{}%
  \ifstrequal{#1}{L}{Machine Learning Symbols}{}%
  \ifstrequal{#1}{M}{Probability Distribution Symbols}{}%
]\vspace{10pt}} 

\title{Detection of Clouds in Multiple Wind Velocity Fields using Ground-based Infrared Sky Images}

\author{
 Guillermo Terr\'en-Serrano \\
  Department of Electrical and Computer Engineering \\
  The University of New Mexico \\
  Albuquerque, NM 87131, United States\\
  \texttt{guillermoterren@unm.edu} \\
 \And
  Manel Mart\'inez-Ram\'on \\
  Department of Electrical and Computer Engineering \\
  The University of New Mexico \\
  Albuquerque, NM 87131, United States\\
  \texttt{manel@unm.edu} \\
}

\begin{document}

\maketitle

\begin{figure}[!htb]
    \centering
    \includegraphics[width=1.\linewidth]{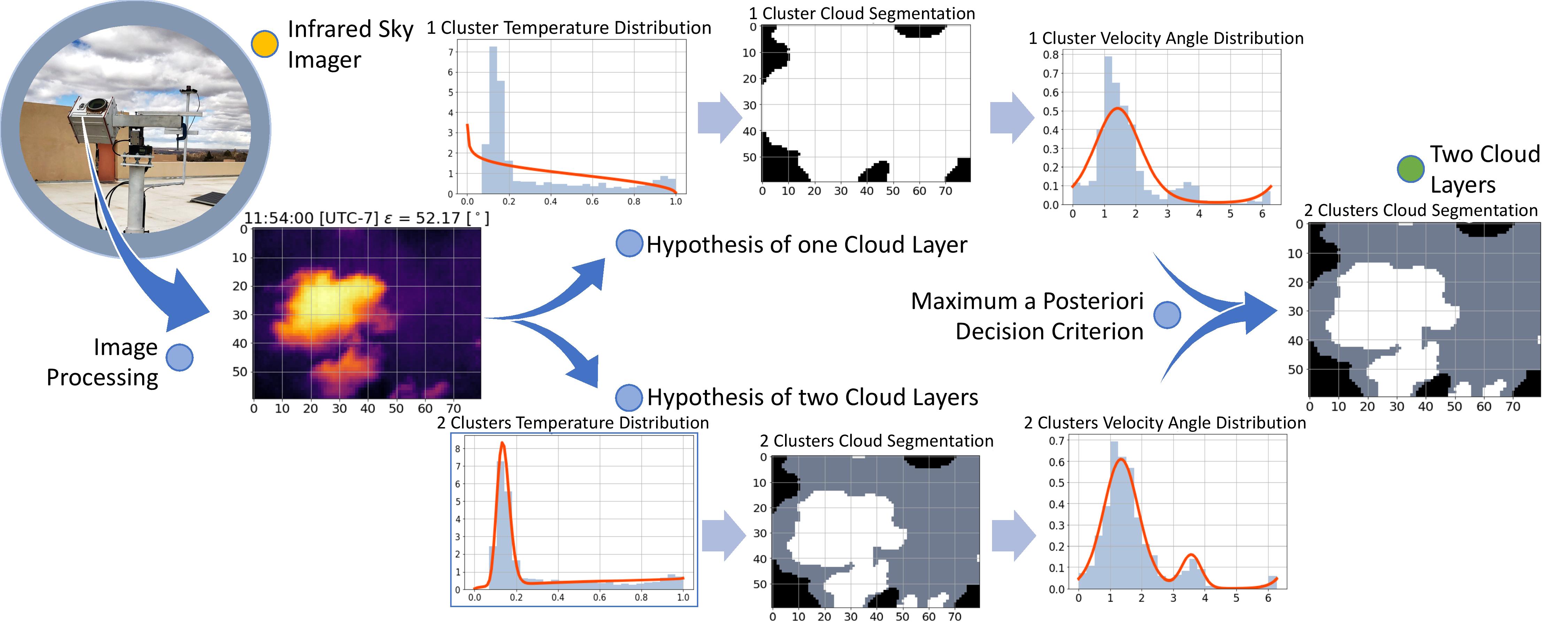}
\end{figure} 
    
\begin{abstract}
    Horizontal atmospheric wind shear causes wind velocity fields to have different directions and speeds. In images of clouds acquired using ground-based sky imagers, clouds may be moving in different wind layers. To increase the performance of an intra-hour global solar irradiance forecasting algorithm, it is important to  detect  multiple layers of clouds. The information provided by a solar forecasting algorithm is necessary to optimize and schedule the solar generation resources and storage devices in a smart grid. This investigation studies the performance of unsupervised learning techniques when detecting the number of cloud layers in infrared sky images. The images are acquired using an innovative infrared sky imager mounted on a solar tracker. Different mixture models are used to infer the distribution of the cloud features. Multiple Bayesian metrics and a sequential hidden Markov model  are implemented to find the optimal number of clusters in the mixture models, and their performances are compared. The motion vectors are computed using a weighted implementation of the Lucas-Kanade algorithm. The correlations between the cloud velocity vectors and temperatures are analyzed to find the method that leads to the most accurate results. We have found that the sequential hidden Markov model outperformed the detection accuracy of the Bayesian metrics.
\end{abstract}


\printnomenclature[2cm]
\nomenclature[A]{\textbf{PV}}{Photovoltaic}
\nomenclature[A]{\textbf{SG}}{Smart Grid}
\nomenclature[A]{\textbf{GSI}}{Global Solar Irradiance}
\nomenclature[A]{\textbf{NWP}}{Numerical Weather Prediction}
\nomenclature[A]{\textbf{IR}}{Infrared}
\nomenclature[A]{\textbf{DAQ}}{Data Acquisition System}
\nomenclature[A]{\textbf{HPC}}{High Performance Computer}
\nomenclature[A]{\textbf{FOV}}{Field of View}
\nomenclature[A]{\textbf{LK}}{Lucas-Kanade}
\nomenclature[A]{\textbf{HMM}}{Hidden Markov Model}
\nomenclature[A]{\textbf{EM}}{Expectation-Maximization}
\nomenclature[A]{\textbf{BeMM}}{Beta Mixture Model}
\nomenclature[A]{\textbf{GaMM}}{Gamma Mixture Model}
\nomenclature[A]{\textbf{GMM}}{Gaussian Mixture Model}
\nomenclature[A]{\textbf{WLK}}{Weighted Lucas-Kanade}
\nomenclature[A]{\textbf{LS}}{Least Squares}
\nomenclature[A]{\textbf{WLS}}{Weighted Least Squares}
\nomenclature[A]{\textbf{VMMM}}{Von Mises Mixture Model}
\nomenclature[A]{\textbf{MAP}}{Maximum a Posteriori}
\nomenclature[A]{\textbf{ML}}{Maximum Likelihood}
\nomenclature[A]{\textbf{BGaMM}}{Bivariate Gamma Mixture Model}
\nomenclature[A]{\textbf{BIC}}{Bayesian Information Criterion}
\nomenclature[A]{\textbf{ICL}}{Integrated Classification Likelihood}
\nomenclature[A]{\textbf{AIC}}{Akaike Information Criterion}
\nomenclature[A]{\textbf{CLC}}{Classification Likelihood Criterion}

\nomenclature[F]{$i$}{Index of the vertical coordinate of a pixel}
\nomenclature[F]{$j$}{Index of the horizontal coordinate of a pixel}
\nomenclature[F]{$t$}{Index of the time or object position on the time}
\nomenclature[F]{$T_{i,j}$}{Temperature of a pixel in Kelvin}
\nomenclature[F]{$\bar{T}_{i,j}$}{Normalized Temperature of a pixel between 0 and 1}
\nomenclature[F]{$\tilde{T}_{i,j}$}{Normalized Temperature of a pixel between 0 and $\infty$}
\nomenclature[F]{$\mathcal{I}$}{Intensity of a object in an image}
\nomenclature[F]{$x$}{Object position on the x direction}
\nomenclature[F]{$y$}{Object position on the y direction}
\nomenclature[F]{$\Delta x$}{Small displacement on the x direction}
\nomenclature[F]{$\Delta y$}{Small displacement on the y direction}
\nomenclature[F]{$\Delta t$}{Small displacement on the time}
\nomenclature[F]{$u$}{Velocity component in the x direction of an object in an image}
\nomenclature[F]{$v$}{Velocity component in the y direction of an object in an image}
\nomenclature[F]{$\mathbf{I}$}{An image}
\nomenclature[F]{$\mathbf{I}_x$}{Finite differentials of an image in the x direction}
\nomenclature[F]{$\mathbf{I}_y$}{Finite differentials of an image in the y direction}
\nomenclature[F]{$\mathbf{I}_t$}{Finite differentials of an image in time}
\nomenclature[F]{$\mathbf{K}_x$}{Differential kernel in the x direction}
\nomenclature[F]{$\mathbf{K}_y$}{Differential kernel in the y direction}
\nomenclature[F]{$\mathbf{K}_t$}{Differential kernel in the time}
\nomenclature[F]{$\sigma$}{Amplitude parameter of the time differential kernel}
\nomenclature[F]{$L$}{Number of cloud layers in an image}
\nomenclature[F]{$l$}{A cloud layer index in an image}
\nomenclature[F]{$w$}{Sliding window width}
\nomenclature[F]{$n$}{Index of the vertical coordinate of pixel in the sliding window}
\nomenclature[F]{$m$}{Index of the horizontal coordinate of pixel in the sliding window}
\nomenclature[F]{$\mathbf{X}_{i,j}$}{Matrix containing the finite differentials in the x an y direction of the pixels within the sliding window in pixel $i,j$}
\nomenclature[F]{$\mathbf{y}_{i,j}$}{Velocity vector in pixel $i,j$}
\nomenclature[F]{$\mathbf{v}_{i,j}$}{Vector containing the finite time differentials of the pixels within the sliding window in pixel $i,j$}
\nomenclature[F]{$\boldsymbol{\Gamma}_{i,j}$}{Diagonal matrix of mixture model posterior probability  within the sliding window in pixel $i,j$}
\nomenclature[F]{$\mathbf{z}_{i,j}$}{Mixture Model latent variable of pixel $i,j$}
\nomenclature[F]{$\tau$}{Coefficient regularization parameter of the WLS}
\nomenclature[F]{$\mathbf{U}$}{Velocity vector x component in all the pixels in an image}
\nomenclature[F]{$\mathbf{V}$}{Velocity vector y component in all the pixels in an image}
\nomenclature[F]{$\mathbf{R}$}{Velocity vector magnitude in all the pixels in an image}
\nomenclature[F]{$\boldsymbol{\Phi}$}{Velocity vector angle in all the pixels in an image}
\nomenclature[F]{$N$}{Number of pixels in the horizontal direction}
\nomenclature[F]{$M$}{Number of pixels in the vertical direction}

\nomenclature[L]{$i$}{Index of an arbitrary sample}
\nomenclature[L]{$t$}{Step in the EM optimization}
\nomenclature[L]{$x_i$}{A feature vector}
\nomenclature[L]{$\mathcal{Q}(\cdot)$}{Expected complete data log-likelihood}
\nomenclature[L]{$\boldsymbol{\pi}$}{Cluster weight in a mixture model}
\nomenclature[L]{$\boldsymbol{\alpha}$}{Dirichlet distribution parameters of the cluster weight prior in a MAP mixture model.}
\nomenclature[L]{$\boldsymbol{\theta}$}{Set of parameter in a mixture model.}
\nomenclature[L]{$\hat{\boldsymbol{\theta}}$}{Optimal set of parameter in a mixture model.}
\nomenclature[L]{$\hat{\boldsymbol{\alpha}}$}{Optimal set of prior parameters.}
\nomenclature[L]{$N$}{Arbitrary number of samples}
\nomenclature[L]{$\lambda$}{Number of parameters in a mixture model}
\nomenclature[L]{$\boldsymbol{\Theta}$}{Set of parameters in a HMM}
\nomenclature[L]{$L_t$}{Number of Cluster in a mixture model and latent variable of the HMM}
\nomenclature[L]{$\psi(\cdot)$}{State of the system function}
\nomenclature[ML]{$\beta$}{State of the system probability parameter}

\nomenclature[M]{$X$}{A random variable}
\nomenclature[M]{$Y$}{A random variable}
\nomenclature[M]{$\alpha$}{Shape parameter of a Gamma, Bivariate Gamma and Beta distribution}
\nomenclature[M]{$\beta$}{Rate parameter of a Gamma, Bivariate Gamma and Beta distribution}
\nomenclature[M]{$a$}{Parameter of a Bivariate Gamma distribution}
\nomenclature[M]{$\kappa$}{Concentration parameter of Von Mises distribution}
\nomenclature[M]{$\mu$}{Mean parameter of a Normal and Von Mises distribution}
\nomenclature[M]{$\Sigma$}{Covariance parameter of a Multivariate Normal Distribution}
\nomenclature[M]{$\Gamma (\cdot)$}{Gamma function}
\nomenclature[M]{$\mathcal{G} (\cdot)$}{Gamma Distribution}
\nomenclature[M]{$\mathcal{BG} (\cdot)$}{Bivariate Gamma Distribution}
\nomenclature[M]{$\mathcal{B} (\cdot)$}{Beta Distribution}
\nomenclature[M]{$\mathcal{VM} (\cdot)$}{Von Mises Distribution}
\nomenclature[M]{$\mathcal{N} (\cdot)$}{Normal Distribution}
\nomenclature[M]{$\Gamma^\prime (\cdot)$}{Derivative of the Gamma function}
\nomenclature[M]{$I_\nu (\cdot)$}{Modified Bessel function of order $\nu$}
\nomenclature[M]{$\mathrm{B}(\cdot)$}{Beta function}
\nomenclature[M]{$\psi (\cdot)$}{Digamma Function}

\section{Introduction}


The ongoing transition toward energy generation systems that produce low-carbon emissions is increasing the penetration of renewable energies in the power grid \cite{ZHAO2020}. However, the only three renewable sources that can produce enough power to fulfill the demand are geothermal, biomass and solar. In particular, solar energy has the potential to become the primary source of power due to its availability and capability \cite{KABIR2018}. A Smart Grid (SG) may optimize the dispatch of energy in large Photovoltaic (PV) power plants to meet demand using recent advances in information and communication technologies \cite{FENG2021}.


The power generated by PV systems is affected by Global Solar Irradiance (GSI) fluctuations that reach the surface of PV panels \cite{OTANI1997, SUN2018}. Shadows projected by moving clouds produce mismatch losses \cite{lappalainen2013}. Although certain configurations of PV arrays reduce the impact of the losses, they still are outside of the allowed range demanded by grid operators \cite{lappalainen2017}. Forecasting of power output will equip a SG powered by PV systems with the technology necessary for regulating the dispatch of energy \cite{MARTIN2010, AHMED2020}. Nevertheless, PV power plants have different physical configurations \cite{QUAN2020}. PV cells and batteries degrade following unique patterns \cite{FIGUEIREDO2019}. For these reasons, the predicted power output cannot be directly extrapolated among the PV systems connected to the same SG \cite{SHIMOSE2014, WOLFF2016, WANG2017, ESEYE2018}. The forecasting of GSI over a grid projected on the Earth's surface facilitates the prediction of power output from PV systems located within the span of the grid \cite{WANG2020, MILLER2018}.


The formation of clouds is a phenomenon restricted by the Tropopause \cite{KALNAY2003}. Different types of clouds are expected to form at different altitudes within the Troposphere \cite{LAMB2011}. The magnitude of the wind velocity field increases with the altitude in the lower atmosphere \cite{WIZELIUS2007, WIZELIUS2012}. The wind gradient may also change its direction due to the friction of the wind with the surface of the Earth. The planetary boundary layer (the lowest part of the Troposphere) \cite{CHARLSON2000} is the point where the wind shear causes low level clouds to move in a different direction and speed of that from high level clouds \cite{bousquet2007, bousquet2008, TERREN2020b}.


Numerical Weather Prediction (NWP) models are computationally expensive for the forecasting resolution necessary in these applications \cite{MATHIESEN2011, PEREZ2013, MATHIESEN2013, VERZIJLBERGH2015, AGUIAR2016, MURATA2018}. GSI forecasting models which include ground weather features from meso-scale meteorology have problems of collinearity \cite{GARCIA2018}. Cloud information extracted from geostationary satellite images improved the performance of solar irradiance forecasting with respect to NWP models \cite{MARCHESONI2020, ALONSO2020}. However, real-time applications of GSI forecasting using satellite imaging are not feasible due to communications delays \cite{JANG2016, MALLIKAI2020}. Ground-based sky imaging systems are an efficient and low-cost alternative for satellite imaging \cite{cervantes2016, richardson2017, KONG2020}. The performances of solar irradiance or PV power output forecasting algorithms are increased when visible and infrared (IR) ground-based sky imaging systems are installed on a solar tracker \cite{CHU2016, MAMMOLI2013}.


Attaching a fish-eye lens to a low-cost visible light camera can provide sky images with large Field of View (FOV) \cite{CALDAS2019, SHAFFERY2020}. The disadvantage of using visible light sky imaging systems in solar forecasting is that the intensity of the pixels in the circumsolar area are saturated \cite{CHOW2011}. Radiometric long-wave IR cameras of uncooled microbolometers are low-cost and widely available \cite{SHAW2013}. These types of cameras have been used to analyze the radiation emitted by gases and clouds in the atmosphere \cite{SHAW2005}. Cloud statistics may be computed \cite{THURAIRAJAH2007} to establish optical links for applications that involve Earth-space communication \cite{NUGENT2009}. The measured temperature of clouds depends on the air temperature on the ground \cite{TERREN2020a}, so the calibration of these cameras is important to perform accurate measurements \cite{NUGENT2013}. Merging thermal images acquired from multiple IR cameras mounted on a dome-shaped plane can provide thermal images with a larger FOV \cite{MAMMOLI2019}.


The wind velocity field shown in a sequence of cloud images is a physical process that is assumed to have a limited complexity \cite{REN2018}. A sequence of IR images allows the derivation of physical features from moving clouds in a wind velocity field. These are more interpretable for modelling physical processes. The features are temperature, velocity vectors, and height \cite{Escrig2013, TERREN2020c}. It has been found that the advantage of using unsupervised learning algorithms is that the response to a sequence of images is expected to depend on the physical process that the images represent rather than the intensity of their pixels \cite{hinton1999}. Unsupervised learning methods, in special mixture models, infer the probability density functions of the observations without prior information \cite{hastie2009}. 


Hidden Markov Models (HMM) were introduced to model linear sequences of discrete latent variables or states as a Markov process. These models are popular in computer vision and pattern recognition applications in which the current state of a system in an image is modelled in terms of previous states \cite{bunke2001}. HMM have been used to detect and analyze temperature distributions of images acquired from IR thermal imaging systems \cite{Bharadwaj2002}. It is possible to apply HMM to model stochastic physical processes \cite{bellone2000, hughes1999} in image classification \cite{li2000}, and object recognition \cite{bashir2007}.


The unsupervised learning algorithm proposed in this investigation does inference over the number of wind velocity fields in a sequence of images using features extracted from the clouds. IR images of clouds are obtained using an innovative Data Acquisition (DAQ) system mounted in a solar tracker \cite{TERREN2020a}. The velocity vectors are computed using a weighted variation of the standard Lucas-Kanade (LK) method \cite{LUCAS1981}. The velocity vector of each pixel is computed in a weighted window of neighboring pixels. The weight of a pixel is the posterior probability of belonging to the lower or the upper cloud layer. The obtained velocity vectors for each cloud layer are averaged together weighting them by the posterior probability of each cloud layer.

A real-time probabilistic model is implemented to detect the number of layers in an IR image \cite{stauffer1999}. The proposed model is an HMM that models the hidden process of the number of wind velocity fields in a sequence of images \cite{chen2007, wei2008}. The motion of the clouds on a sequence of images is used to calculate the velocity vectors. The temperature and height of the pixels are also extracted from the cloud images. The distributions of the features are inferred with different parametric mixture models \cite{mclachlan1988, mclachlan2004}. The mixture models are optimized using the Expectation-Maximization (EM) algorithm \cite{BISHOP2006, MURPHY2012}. The label switching of the mixture models is solved using the average height of each distribution in the mixture model \cite{stephens2000}.

\section{Methodology}

The long-wave IR camera provides an uniform thermal image applying the Wein's displacement law to the radiation emitted by a black body. Wein's displacement law says that emitted radiation is inversely proportional to the temperature,  the black body radiation maxima is at different wavelengths depending on the temperature. In our application, the feasible cloud temperatures are within the long-wave infrared spectrum \cite{GHOSH1994}.

A pixel of the camera frame is defined by a pair of euclidean coordinates $\mathbf{X} = \{ (i, j ) \mid \forall i = 1, \ldots, M, \ \forall j = 1, \ldots, N \}$, and the temperature of each one of the pixels is defined in Kelvin degrees as $ \mathbf{T}^{(t)} = \{ {T}{}^{(t)}_{i,j} \in \mathbb{R} \mid \forall i = 1, \ldots, M, \ \forall j = 1, \ldots, N \}$, where $t$ represents a process defined as $t \in \left( 0, \infty \right]$, that is a sequence of IR images ordered chronologically. 

When there are multiple layers of clouds in an image, a mixture model is expected to have multiple clusters. In order to infer the distribution of the temperatures or heights using a Beta Mixture Model (BeMM), the features are first normalized to the domain of a beta distribution $\bar{T}_{i,j} = [ T^{(t)}_{i,j} - \mathrm{min} ( \mathbf{T}^{(t)} ) ] / [\mathrm{max} ( \mathbf{T}^{(t)} ) - \mathrm{min} ( \mathbf{T}^{(t)} ) ]$. When the inference is performed with a Gamma Mixture Model (GaMM), the temperatures are normalized to the domain of the gamma distribution $\tilde{T}_{i,j} = T^{(t)}_{i,j} - \mathrm{min} ( \mathbf{T}{}^{(t)} ) $. The heights (in kilometers) are within the domain of the gamma distribution. When the inference is performed using a Gaussian Mixture Model (GMM), the temperatures and heights do not require normalization.

\subsection{Weighted Lucas-Kanade}

In current computer vision literature, there are three primary methods to estimate the motion of objects in a sequence of images: Lucas-Kanade \cite{LUCAS1981}, Horn-Schunck \cite{HORN1981} and Farneb{\"a}ck \cite{FARNEBACK2003} methods. These three methods are based on the space-time partial derivatives between two consecutive frames. The techniques to estimate the motion vectors in an image are sensitive to the intensity gradient of the pixels. An atmospheric model is implemented to remove the gradient produced by the Sun's direct irradiance and the atmospheric scattered irradiance (both of which routinely appear on the images in the course of the year). A persistent model of the outdoor germanium window of the IR camera removes debris and water spots that appear in the image \cite{TERREN2021a}. In this investigation, it is implemented a Weighted Lucas-Kanade (WLK) method.

\subsubsection{Optical Flow}

The optical flow equation considers that exists a small displacement $\Delta x$ and $\Delta y$ in the direction of an object in an image. The object is assumed to have constant intensity $\mathcal{I}$ between two consecutive frames. The frames are separated in time by small time increment $\Delta t$,
\begin{equation}
    \mathcal{I} \left( x, y, t \right) = \mathcal{I} \left( x + \Delta x, y + \Delta y, t + \Delta t \right).
\end{equation}
Assuming that the difference in intensity between neighboring pixels is smooth and that brightness of a pixel in consecutive frames is constant, the Taylor series expansion is applied and following equation obtained,
\begin{equation}
    \mathcal{I} \left( x + \Delta x, y + \Delta y, t + \Delta t \right) = \mathcal{I}  \left( x, y, t \right) + \frac{\partial \mathcal{I} }{\partial x} \Delta x + \frac{\partial \mathcal{I} }{\partial y} \Delta y + \frac{\partial \mathcal{I} }{\partial t} \Delta t.
\end{equation}
The factors are simplified combining the last two equation,
\begin{equation}
    \frac{\partial \mathcal{I} }{\partial x} \Delta x + \frac{\partial \mathcal{I} }{\partial y} \Delta y + \frac{\partial \mathcal{I} }{\partial t} \Delta t = 0.
\end{equation}
The velocity of an object is derived dividing the terms of the displacement by the increment of time $\Delta t$, 
\begin{equation}
    \frac{\partial \mathcal{I} }{\partial x} \frac{\Delta x}{\Delta t} + \frac{\partial \mathcal{I} }{\partial y} \frac{\Delta y}{\Delta t} + \frac{\partial \mathcal{I} }{\partial t} \frac{\Delta t}{\Delta t} = 0.
\end{equation}
The velocity components are defined as $u$ and $v$ so that,
\begin{equation}
    \frac{\partial \mathcal{I} }{\partial x} u + \frac{\partial \mathcal{I}}{\partial y} v + \frac{\partial \mathcal{I} }{\partial t} = 0.
\end{equation}
This equation is known as the aperture problem,
\begin{equation}
    \mathcal{I}_x u + \mathcal{I}_y v = - \mathcal{I}_t,
\end{equation}
where $\mathcal{I}_x = \partial \mathcal{I} / \partial x$, $\mathcal{I}_y = \partial \mathcal{I} / \partial y $ and $\mathcal{I}_t = \partial \mathcal{I}/ \partial t $  are the derivatives for notation simplification. 

The 2-dimensional derivatives are approximated using convolutional filters in a image \cite{HAST2014}. Let us define a discrete time and space sequence of images as  $\mathbf{I}^{(t)} = \{ \mathcal{I}^{(t)}_{i,j} \in \mathbb{R}^{[0, 2^8)} \mid i = 1, \dots, N, \ j = 1, \dots, M \}$. The finite differences method is applied to compute the derivatives,
\begin{equation}\label{eq:finite_differences}
    \begin{split}
        \mathbf{I}_x^\prime &= \mathbf{I}^{(t - 1)} \star \mathbf{K}_x \\
        \mathbf{I}_y^\prime &= \mathbf{I}^{(t - 1)} \star \mathbf{K}_y \\
        \mathbf{I}_t^\prime &= \mathbf{I}^{(t - 1)} \star \mathbf{K}_t  + \mathbf{I}^{(t)} \star -  \mathbf{K}_t
    \end{split}
\end{equation}
where $\star$ represent a 2-dimensional convolution, and $\mathbf{I}^{(t - 1)}$ and $\mathbf{I}^{(t)}$ are the first and second consecutive frames. $\mathbf{K}_x$, $\mathbf{K}_y$ and $\mathbf{K}_t$ are the differential kernels in the x, y and t direction respectively, 
\begin{equation}
    \mathbf{K}_x = \begin{bmatrix}
         -1 & 1 \\
         -1 & 1
    \end{bmatrix}, \quad
    \mathbf{K}_y = \begin{bmatrix}
         -1 & -1 \\
         1 & 1
    \end{bmatrix}, \quad
    \mathbf{K}_t = \sigma \begin{bmatrix}
         1 & 1 \\
         1 & 1
    \end{bmatrix},
\end{equation}
the parameter $\sigma$ is the amplitude of the temporal kernel. This parameter may be cross-validated when the velocity field is known.

\subsubsection{Lucas-Kanade}

The LK method proposes to find the solution for the optical flow equations via Least Squares (LS). The optical flow equation is solved using a local image of the pixels within a sliding window. In this research, the LK method is extended for multiple importance weights. A sliding window is defined with odd width $W = 2w+1$, where $w$ is the window size parameter, which has to be cross-validated.

Now we assume that the image may contain more than one velocity field. Then, at pixel $i,j$ of layer $l$, we define $1 \leq l \leq L$ hypothesis over the possible velocities $\mathbf{v}^{(l)}_{i,j}$. 

If the position of the central pixel of the window is defined as $i,j$, then the dependent and independent variables can be defined as 
\begin{equation}
    \begin{split}
        \mathbf{x}_{i-m,j-n} &=
            \begin{bmatrix}
                \mathbf{I}^\prime_{x} \left(i-m, j-n \right) \\ 
                \mathbf{I}^\prime_{y} \left(i-m, j-n \right)
            \end{bmatrix}, \quad
        \mathbf{v}^{(l)}_{i, j} =
            \begin{bmatrix}
                u^{(l)}_{i} \\ 
                v^{(l)}_{j}
            \end{bmatrix}, \ 
        \\
        y_{i-m, j-n} &= -\mathbf{I}^\prime_{t} \left( i-m, j-n \right), \quad
        - w \leq m \leq  w, \quad - w \leq n \leq  w.
    \end{split}
\end{equation}
Each velocity vector is associated to a posterior probability $\gamma_{i,j}^{(l)} \triangleq  p ( z_{i,j} = l \mid T_{i,j})$ where $z_{i,j} = l$ is a latent variable that indicates that the velocity field at pixel $i,j$ is corresponds to cloud layer $l$. These posteriors will be estimated in the next section.

We introduce a Weighted Least Squares (WLS) approach \cite{SIMON2003}, whose weights are posterior probabilities $\gamma^{(l)}_{i,j}$. Assume an extended vector ${\bf y}_{i,j}$ containing all instances of  ${\bf y}_{i-m,j-n}$ and a matrix ${\bf X}_{i,j}$ containing all ${\bf x}_{i-m,j-n}$.

Instead of minimizing the mean square error, we can maximize the expectation of the unnormalized log-posterior 
\begin{equation}\label{eq:log_posterior}
    \begin{split}
    &\mathbb{E} \left\{\log \left[ p \left({\bf y}_{i,j} \middle| {\bf X}_{i, j},{\bf v}^{(l)}_{i,j} \right) p \left( {\bf v}^{(l)}_{i,j} \right) \right] \right\} = \\
    &= \mathbb{E} \left\{ \log \left[ \prod_{m,n,l} \left[ p \left( {\bf y}_{i-m,j-n} \middle| {\bf x}_{i-m,j-n}, {\bf v}_{i,j}^{(l)} \right) \right]^{\mathbb{I} \left( z_{i,j} = l \right) } p \left({\bf v}^{(l)}_{i,j} \right) \right] \right\} \\
    &= \sum_{m,n,l} \mathbb{E} \left[ \mathbb{I} \left( z_{i-m, j-n} = l \right)\right] \log p \left({\bf y}_{i-m, j-n} \middle| {\bf x}_{i-m, j-n}, {\bf v}_{i,j}^{(l)} \right)  + \log p \left({\bf v}^{(l)}_{i,j} \right) \\
    & = \sum_{m,n,l} \gamma^{(l)}_{i - m, j - n} \log p \left( {\bf y}_{i-m, j-n, l} \middle| {\bf x}_{i-m,j-n},{\bf v}^{(l)}_{i,j} \right) + \log p({\bf v}_{i,j}^{(l)}) \\
    &= - \sum_{m, n, l} \gamma^{(l)}_{i-m,j-n} \| {\bf v}^{(l)\top}_{i, j} {\bf x}_{i - m, j - n} - {\bf y}_{i - m, j - n}\|^2 - \tau \| {\bf v}^{(l)}_{i, j} \|^2 + \text{constant}
    \end{split}
\end{equation}
where $\gamma^{(l)}_{i - m, j - n} = \mathbb{E}[ \mathbb{I}(z_{i - m, j - n} = l)]$ is the posterior probability of the velocity field, $\mathbb{I}(\cdot)$ is the indicator function, and where we assumed that both probabilities are Gaussian distributions, where the first one is a multivariate Gaussian modelling the error, which has a given variance $\sigma_n^2$, and the second one is a multivariate Gaussian modelling the prior over the velocities, whose covariance is an identity $\mathbf{I}_{2 \times 2}$. Thus, $\tau = \sigma_n^{2}$ plays the role of a regularization parameter, and it may be validated or inferred by maximizing the likelihood term \cite{sundararajan2000}.

By computing the gradient of the expression with respect to ${\bf v}^{(l)}_{i,j}$ and nulling it, we obtain the solution,
\begin{equation}
    \mathbf{v}^{(l)}_{i,j} = \left( \mathbf{X}_{i,j} \boldsymbol{\Gamma}_{i,j}^{(l)} \mathbf{X}_{i,j}^\top + \tau \mathbf{I}_{2L \times 2L} \right)^{-1} \mathbf{X}_{i,j} \boldsymbol{\Gamma}_{i,j}^{(l)} \mathbf{y}_{i,j}
\end{equation}
where ${\boldsymbol \Gamma}_{i,j}^{(l)}$ is a diagonal matrix containing all posteriors of the $l$ cluster. 

The estimated velocity components are defined as $\mathbf{U}{}^{(l)} = \{ u{}^{(l)}_{i,j} \in \mathbb{R} \mid i = 1, \dots, N, \ j = 1, \dots, M \}$ and $\mathbf{V}{}^{(l)}= \{ v^{(l)}_{i,j} \in \mathbb{R} \mid i = 1, \dots, N, \ j = 1, \dots, M \}$.

The obtained velocity components for each posterior $l$, are averaged weighting the vectors components by their posterior,
\begin{equation}
    \mathbf{U} = \sum_{l = 1}^L \left( \boldsymbol{\Gamma}^{(l)} \odot \mathbf{U}^{(l)} \right); \quad \mathbf{V} = \sum_{l = 1}^L \left( \boldsymbol{\Gamma}^{(l)} \odot \mathbf{V}^{(l)} \right),
\end{equation}
where $\odot$ is the Hadamard product. The result are the velocity components for each pair of coordinates in the original frame $\mathbf{U}, \mathbf{V} \in \mathbb{R}^{N \times M}$. The velocity vectors defined in polar coordinates have magnitude $\mathbf{R} = ( \mathbf{U} \odot \mathbf{U} + \mathbf{V} \odot \mathbf{V})^{1/2}$ and angle $\boldsymbol{\Phi} = \mathbf{arctan2} ( \mathbf{U}, \mathbf{V} )$.

\subsection{Maximum a Posteriori Mixture Model}

When the clouds are moving in multiple wind velocity fields the distributions of the temperatures, heights and velocity vectors components are expected to be distributed in multiple clusters in the feature space of the observation.

Mixture models are implemented to infer the distributions of the physical features extracted from IR cloud images. These physical features have different domain so the inference is implemented using probability functions defined in each one of these domains. Thus, the distribution of the velocity vectors defined in Cartesian coordinates is inferred with a multivariate GMM. The inference of the velocity vectors when they are defined in polar coordinates, is performed independently for each component using a GaMM and Von Mises Mixture Model (VMMM) for the magnitude and the angle respectively.

In this section, we propose the use of different mixture models to infer probability functions that approximate better the actual distribution of a features with the aim of detecting the most likely number of wind velocity fields on an image and their pixelwise posterior probabilities $\gamma^{(l)}_{i,j}$. The formulation of the proposed mixture models includes a prior distribution on the cluster weights in order to avoid overfitting.

\subsubsection{Expectation-Maximization}

Let us consider that $\mathbf{x}_i$ are observations (i.e. feature vectors) which we wish to model as a mixture model, and that $z_i$ are the corresponding latent variables of their cluster index. The optimal set of parameters in a mixture model can be computed using the EM algorithm. The implementation of the EM algorithm guarantees a smooth convergence to a local maximum following an iterative approach consisting of two steps \cite{BISHOP2006}. The maximized function is the expected complete data log-likelihood plus the log-prior,
\begin{equation}\label{eq:CDLL}
    \begin{split}
        \mathcal{Q} \left( \boldsymbol{\theta}, \boldsymbol{\theta}_{t - 1} \right) &\triangleq \mathbb{E} \left[ \sum_{i = 1}^N  \log p \left( x_i, z_i \middle| \boldsymbol{\theta} \right) + \log p \left( \boldsymbol{\pi} \middle| \boldsymbol{\alpha} \right) \right] \\
        &= \sum_{i = 1}^N \mathbb{E} \left[ \log \left( \prod^{L}_{l = 1} \left[ \pi^{(l)} p \left( x_i \middle| \boldsymbol{\theta}^{(l)} \right)^{\mathbb{I} \left( z_i = l \right)} \right] \right) \right] + \log p \left( \boldsymbol{\pi} \middle| \boldsymbol{\alpha} \right) \\
        &= \sum_{i = 1}^N \sum_{l = 1}^L p \left( z_i = l \middle| x_i, \boldsymbol{\theta}^{(l)}_{t - 1} \right) \log \left[ \pi^{(l)} p \left( x_i \middle| \boldsymbol{\theta}^{(l)} \right) \right] + \log p \left( \boldsymbol{\pi} \middle| \boldsymbol{\alpha} \right) \\
        &= \sum_{i = 1}^N \sum_{l = 1}^L \gamma_i^{(l)} \log \left[ \pi^{(l)} p \left( x_i \middle| \boldsymbol{\theta}^{(l)} \right) \right] + \log p \left( \boldsymbol{\pi} \middle| \boldsymbol{\alpha} \right),
    \end{split}
\end{equation}
where $t$ represents the iteration of the algorithm, and the posterior probability introduced in Eq. \eqref{eq:log_posterior} appears here as 
\begin{equation}
    \gamma^{(l)}_i \triangleq p \left( z_i = l \middle| x_i, \boldsymbol{\theta}^{(l)}_{t - 1},{\boldsymbol \alpha} \right),
\end{equation}
which is commonly referred to as the responsability of cluster $l$ in  sample $i$. A prior $p \left( \boldsymbol{\pi} \middle| \boldsymbol{\alpha} \right)$ with parameters $\boldsymbol \alpha$ is introduced for probabilities $\boldsymbol \pi$.
The initialization of the EM starts by randomly assigning a set of parameters and a prior. In the expectation step of the EM algorithm, a posterior $\gamma_i^{(l)}$ is assigned to each sample using the likelihood function, 
\begin{equation}
    \gamma^{(l)}_i = \frac{\pi^{(l)} p \left( x_i \middle| \boldsymbol{\theta}^{(l)}_{t - 1} \right)}{\sum_{l = 1}^L \pi^{(l)} p \left( x_i \middle| \boldsymbol{\theta}^{(l)}_{t - 1} \right)}.
\end{equation}

In the maximization step, the parameters that maximize the complete data log-likelihood plus the log-prior are found analytically, as it will be shown further.
The prior $p(\boldsymbol{\pi} | \boldsymbol{\alpha})$ is a Dirichlet distribution $\boldsymbol{\pi} \sim \mathrm{Dir} (\boldsymbol{\alpha})$, and $\alpha^{(l)} \geq 1$. The mixture weights are updated using the posterior probabilities \cite{MURPHY2012},
\begin{equation}
    \pi^{(l)} = \frac{\alpha^{(l)} - 1 + \sum^N_{i = 1} \gamma^{(l)}_i}{N - L + \sum^L_{l = 1} \alpha^{(l)}},
    \label{eq:mm_prior}
\end{equation}
when $\alpha^{(l)} = 1$, the prior is noninformative $p ( \boldsymbol{\pi} | \boldsymbol{\alpha} ) = 0$ and thus the Maximum A Posterior (MAP) estimation is equivalent to the Maximum Likelihood (ML), $\pi^{(l)} = [\sum^N_{i = 1} \gamma_i^{(l)}]/ N$.

The E and M steps are repeated until the complete data log-likelihood have converged to a local maxima. In the case of a quadratic loss function this problem has analytical solution, for instance in a GMM \cite{MURPHY2012}. When the loss function has not analytical solution, it can be solved implementing a numerical optimization method based on the gradient descent.

\subsubsection{Gamma Mixture Model}

The distribution of the magnitude of velocity vectors or the heights can be approximate by mixture of Gamma distributions $X \sim \mathcal{G} ( \alpha^{(l)}, \beta^{(l)} )$ which density function is,
\begin{equation}
    f \left(x_i ; \alpha^{(l)}, \beta^{(l)} \right) = \frac{x_i^{\alpha^{(l)} - 1} e^{- \frac{x_i}{\beta^{(l)}}}}{\beta^{(l)^{\alpha^{(l)}}} \Gamma \left( \alpha^{(l)} \right)} \quad x_i > 0, \quad \alpha^{(l)}, \beta^{(l)} > 0, 
\end{equation}
where $\Gamma ( \alpha^{(l)})$ is the Gamma function.

The log-likelihood of the Gamma density function needed to compute the expected complete data log-likelihood in a GaMM is,
\begin{equation}
    \log p \left( x_i \middle| \alpha^{(l)}, \beta^{(l)} \right) = \left( \alpha^{(l)} - 1 \right) \log x_i - \frac{x_i}{\beta^{(l)}} - \alpha^{(l)} \log \beta^{(l)} - \log \Gamma \left( \alpha^{(l)} \right).
\end{equation}

The maximization step has to be solved via numerical optimization. The gradient w.r.t. $\alpha^{(l)}$ is,
\begin{equation}
    \begin{split}
        \frac{\partial \mathcal{Q} \left( \boldsymbol{\theta}^{(l)} \right)}{\partial \alpha^{(l)}} &= \sum_{l = 1}^L \sum_{i = 1}^N \gamma_i^{(l)} \frac{\partial}{\partial \alpha^{(l)}} \log p \left( x_i \middle| \alpha^{(l)}, \beta^{(l)} \right) \\
        &= \sum_{l = 1}^L \sum_{i = 1}^N \gamma_i^{(l)} \left[ \log x_i - \log \beta^{(l)} - \frac{\Gamma^\prime \left( \alpha^{(l)} \right)}{\Gamma \left( \alpha^{(l)} \right)}\right],
    \end{split} 
\end{equation} 
where $\Gamma^\prime ( \alpha^{(l)} )$ is the derivative of the Gamma function. The gradient w.r.t. $\beta^{(l)}$ is,
\begin{equation}
    \begin{split}
        \frac{\partial \mathcal{Q} \left( \boldsymbol{\theta}^{(l)} \right)}{\partial \beta^{(l)}} &= \sum_{l = 1}^L \sum_{i = 1}^N \gamma_i^{(l)}\left[
        \frac{\partial}{\partial \beta^{(l)}} \log p \left( x_i \middle| \alpha^{(l)}, \beta^{(l)} \right) \right] \\
        &= \sum_{k = 1}^L \sum_{i = 1}^N \gamma_i^{(l)} \left[ \frac{1}{\beta^{(l)}} \left( \frac{x_i}{\beta^{(l)}} - \alpha^{(l)} \right) \right].
    \end{split} 
\end{equation} 

The generalizations of the Gamma distribution for multiple dimensions do not have an unified expression. In fact, the multivariate Gamma distribution is unknown in the exponential family \cite{SEN2014}.

\subsubsection{Bivariate Gamma Mixture Model}

The distribution of magnitude of velocity vectors and heights can be approximated by mixture of bivariate Gamma distributions $X,Y \sim \mathcal{BG} ( \alpha^{(l)}, \beta^{(l)}, a^{(l)} )$ which density function is \cite{SEN2014},
\begin{equation}
    f \left( x_i, y_i ; \alpha^{(l)}, \beta^{(l)}, a^{(l)} \right) = \frac{\beta^{(l)^{\alpha^{(l)}}} x_i^{ \alpha^{(l)} + a^{(l)} - 1} y_i^{ \alpha^{(l)} - 1}}{\Gamma \left( a^{(l)} \right) \Gamma \left( \alpha^{(l)} \right)} e^{- \beta^{(l)} x_i} e^{-x_i y_i} \quad x_i, y_i > 0, 
\end{equation}
where $\Gamma ( \alpha^{(l)})$ is the Gamma function, and the parameters are $\alpha^{(l)}, \beta^{(l)}, a^{(l)} > 0$.

The log-likelihood of the bivariate Gamma density function needed for computing the expected complete data log-likelihood in a Bivariate Gamma Mixture Model (BGaMM) is,
\begin{equation}
    \begin{split}
        \log p & \left( x_i, y_i \middle| \alpha^{(l)}, \beta^{(l)},  a^{(l)} \right) = \alpha^{(l)} \log \beta^{(l)} + \left( \alpha^{(l)} + a^{(l)} - 1 \right) \log x_i + \dots \\
        &\dots + \left( a^{(l)} - 1 \right) \log y_i - \beta^{(l)} x_i - x_i y_i - \log \Gamma \left( \alpha^{(l)} \right) - \log \Gamma \left( a^{(l)} \right).
    \end{split}
\end{equation}

As the maximization of Eq. \eqref{eq:CDLL} has not analytical solution when the likelihood is a bivariate Gamma, the maximization step is solved by numerical optimization. The gradient w.r.t. $\alpha^{(l)}$ is,
\begin{equation}
    \begin{split}
        \frac{\partial \mathcal{Q} \left( \boldsymbol{\theta}^{(l)} \right)}{\partial \alpha^{(l)}} &= \sum_{l = 1}^L \sum_{i = 1}^N \gamma_i^{(l)} \frac{\partial}{\partial \alpha^{(l)}} \log p \left( x_i, y_i \middle| \alpha^{(l)}, \beta^{(l)}, a^{(l)} \right) \\
        &= \sum_{l = 1}^L \sum_{i = 1}^N \gamma_i^{(l)} \left[ \log \beta^{(l)} + \log x_i - \frac{\Gamma^\prime \left( \alpha^{(l)} \right)}{\Gamma \left( \alpha^{(l)} \right)} \right].
    \end{split}
\end{equation}
The gradient w.r.t. $\beta^{(l)}$ is,
\begin{equation}
    \begin{split}
        \frac{\partial \mathcal{Q} \left( \boldsymbol{\theta}^{(l)} \right)}{\partial \beta^{(l)}} &= \sum_{l = 1}^L \sum_{i = 1}^N \gamma_i^{(l)} \left[
        \frac{\partial}{\partial \beta^{(l)}} \log p \left( x_i \middle| \alpha^{(l)}, \beta^{(l)}, a^{(l)} \right) \right] \\
        &= \sum_{l = 1}^L \sum_{i = 1}^N \gamma_i^{(l)} \left[  \frac{\alpha^{(l)}}{\beta^{(l)}} - x_i \right].
    \end{split}
\end{equation}
The gradient w.r.t. $a^{(l)}$ is,
\begin{equation}
    \begin{split}
        \frac{\partial \mathcal{Q} \left( \boldsymbol{\theta}^{(l)} \right)}{\partial a^{(l)}} &= \sum_{l = 1}^L \sum_{i = 1}^N \gamma_i^{(l)} \left[
        \frac{\partial}{\partial \beta^{(l)}} \log p \left( x_i, y_i \middle| \alpha^{(l)}, \beta^{(l)}, a^{(l)} \right) \right] \\
        &= \sum_{l = 1}^L \sum_{i = 1}^N \gamma_i^{(l)} \left[ \log x_i + \log y_i - \frac{\Gamma^\prime \left( a^{(l)} \right)}{\Gamma \left( a^{(l)} \right)} \right].
    \end{split}
\end{equation}

Applying the independence assumption to each component of the Gamma model, the general form of joint density for a multivariate Gamma distribution can be derived, but it needs to be assumed that marginal density functions for each one of random variables are available \cite{SEN2014}.

\subsubsection{Von Mises Mixture Model}

The angular component of the velocity vectors is approximated by a Von Mises distribution $X \sim \mathcal{VM} ( \mu^{(l)}, \kappa^{(l)} )$. The density function of this distribution is,
\begin{equation}
    f \left(x_i ; \mu^{(l)}, \kappa^{(l)} \right) = \frac{e^{\kappa^{(l)} \cos \left( x_i - \mu_l \right) }}{2 \pi I_0 \left( \kappa^{(l)} \right)}, \quad x_i, \mu^{(l)} \in \left[ -\pi, \pi \right], \quad \kappa^{(l)} > 0,
\end{equation}
where $I_0$ represents the modified Bessel function of order 0 that has this formula,
\begin{equation}
    I_\nu \left( \kappa^{(l)} \right) = \left( \frac{\kappa^{(l)}}{2} \right)^\nu \sum_{n = 0}^{\infty} \frac{\left( \frac{1}{4} \left( \kappa^{(l)} \right)^2 \right)^n}{n ! \Gamma( \nu + n + 1)}.
\end{equation}

In the case of mixture of a Von Mises distribution, the data log-likelihood for each cluster is,
\begin{equation}
    \log p \left( x_i \middle| \mu^{(l)}, \kappa^{(l)} \right) = \kappa^{(l)} \cos \left( x_i - \mu^{(l)} \right) - \log 2\pi - \log I_0 \left( \kappa^{(l)} \right).
\end{equation}

The maximization step is solved computing the gradient w.r.t. $\mu^{(l)}$,
\begin{equation}
    \begin{split}
        \frac{\partial \mathcal{Q} \left( \boldsymbol{\theta}^{(l)} \right) }{\partial \mu^{(l)}} &= \sum_{l = 1}^L \sum_{i = 1}^N \gamma_i^{(l)} \left[ \frac{\partial}{\partial \mu^{(l)}}  \log p \left( x_i \middle| \mu^{(l)}, \kappa^{(l)} \right)  \right] \\
        &= \sum_{l = 1}^L \sum_{i = 1}^N \gamma_i^{(l)} \left[ \kappa^{(l)} \sin \left( x_i - \mu^{(l)} \right) \right],
    \end{split}
\end{equation} 
and the gradient w.r.t. $\kappa^{(l)}$,
\begin{equation}
    \begin{split}
        \frac{\partial \mathcal{Q} \left( \boldsymbol{\theta}^{(l)} \right)}{\partial \kappa^{(l)}} &= \sum_{l = 1}^L \sum_{i = 1}^N \gamma_i^{(l)} \left[ \frac{\partial}{\partial \kappa^{(l)}} \log p \left( x_i \middle| \mu^{(l)}, \kappa^{(l)} \right)  \right] \\
        &= \sum_{l = 1}^L \sum_{i = 1}^N \gamma_i^{(l)} \left[ \cos \left( x_i - \mu^{(l)} \right) - \frac{I_1 \left( \kappa^{(l)} \right)}{I_0 \left( \kappa^{(l)} \right)} \right],
    \end{split}
\end{equation} 
where the Bessel function of order 1 is obtained from $\partial I_0 ( \kappa ) / \partial \kappa = I_1 ( \kappa )$.

An extension for the multivariate Von Mises distribution can be found in \cite{NAVARRO2017}, other solutions to the VMMM problem are \cite{SIDDHARTH2014, BANERJEE2005}.

\subsubsection{Beta Mixture Model}

The distribution of the normalized temperatures or heights can be approximated by mixture of beta distributions $ X \sim \mathcal{B} ( \alpha^{(l)}, \beta^{(l)} )$ which density function is,
\begin{equation}\label{likelihood}
    f \left( x_i ; \alpha^{(l)}, \beta^{(l)} \right) = \frac{1}{\mathrm{B} \left( \alpha^{(l)}, \beta^{(l)} \right)} x_i^{\alpha^{(l)} - 1} \left( 1 - x_i \right)^{\beta^{(l)} - 1}, \quad \alpha^{(l)}, \beta^{(l)} > 0,
\end{equation}
where $x_i \in (0, 1)$,  beta function is $\mathrm{B} ( \alpha^{(l)}, \beta^{(l)} ) = [ \Gamma ( \alpha^{(l)} ) \Gamma ( \beta^{(l)} ) ] / [\Gamma ( \alpha^{(l)} + \beta^{(l)} ) ] $, and $\Gamma ( \alpha^{(l)} )$ is the Gamma function.

The log-likelihood of the beta density function, that is needed to compute the expected complete data log-likelihood in the mixture model is,
\begin{equation}
    \log p \left( x_i \middle| \alpha^{(l)}, \beta^{(l)} \right) = \left( \alpha^{(l)} - 1 \right) \log x_i + \left( \beta^{(l)} - 1 \right) \log \left( 1 - x_i \right)- \log \mathrm{B} \left( \alpha^{(l)}, \beta^{(l)} \right).
\end{equation}

The maximization step has to be solved by gradient descent. The gradient w.r.t. $\alpha^{(l)}$ is,
\begin{equation}
    \begin{split}
        \frac{\partial {Q} \left( \boldsymbol{\theta}^{(l)} \right)}{\partial \alpha^{(l)}} &= \sum_{l = 1}^L \sum_{i = 1}^N \gamma_i^{(l)} \frac{\partial}{\partial \alpha^{(l)}} \log p \left( x_i \middle| \alpha^{(l)}, \beta^{(l)} \right) \\
        &= \sum_{i = 1}^N \sum_{l = 1}^L \gamma_i^{(l)} \left[ \log x_i - \psi \left( \alpha^{(l)} \right) + \psi \left( \alpha^{(l)} + \beta^{(l)} \right) \right],
    \end{split}
\end{equation} 
where $\partial \mathrm{B} ( \alpha^{(l)}, \beta^{(l)} ) / \partial \alpha^{(l)} = \mathrm{B} ( \alpha^{(l)}, \beta^{(l)} ) [ \psi ( \alpha^{(l)} ) - \psi ( \alpha^{(l)} + \beta^{(l)} ) ]$, and $\psi ( \cdot )$ is the digamma function, which is $\psi ( \alpha^{(l)} ) = \Gamma^\prime ( \alpha^{(l)} ) / \Gamma ( \alpha^{(l)} )$. The gradient w.r.t. $\beta^{(l)}$ is,
\begin{equation}
    \begin{split}
        \frac{\partial \mathcal{Q} \left( \boldsymbol{\theta}^{(l)} \right)}{\partial \beta^{(l)}} &= \sum_{l = 1}^L \sum_{i = 1}^N \gamma_i^{(l)} \left[
        \frac{\partial}{\partial \beta^{(l)} } \log p \left( x_i \middle| \alpha^{(l)}, \beta^{(l)} \right) \right] \\
        &= \sum_{l = 1}^L \sum_{i = 1}^N \gamma_i^{(l)} \left[ \log \left( - x_i \right) - \psi \left( \beta^{(l)} \right) + \psi \left( \alpha^{(l)} + \beta^{(l)} \right) \right].
    \end{split}
\end{equation} 

In previous work carried out in the implementation of a BeMM clustering, was found that is not optimal to assign the number of clusters in these models applying Bayesian Information Criterion (BIC) \cite{JI2005} (see Subsection \ref{subsec:BIC}). The authors proposed to implement Integrated Classification Likelihood (ICL) instead.

\subsubsection{Gaussian Mixture Model}

The distribution of the velocity components in a Cartesian coordinates system can be approximate by a mixture of multivariate normal distributions $X \sim \mathcal{N} ( \boldsymbol{\mu}^{(l)}, \boldsymbol{\Sigma}^{(l)} )$. The multivariate normal likelihood is,
\begin{equation}\label{EM_likelihood}
    f \left( \mathbf{x} ; \boldsymbol{\mu}^{(l)}, \boldsymbol{\Sigma}^{(l)} \right) = \frac{1}{\sqrt{ \left(2 \pi\right)^d \left| \boldsymbol{\Sigma}^{(l)} \right| }} \cdot \exp \left\{ - \frac{1}{2} \left( \mathbf{x} - \boldsymbol{\mu}^{(l)} \right)^\top \boldsymbol{\Sigma}^{(l)^{-1}} \left( \mathbf{x} - \boldsymbol{\mu}^{(l)} \right) \right\}.
\end{equation}

The log-likelihood of the multivariate density function \cite{MURPHY2012} for computing the expected complete data log-likelihood in the GMM is,
\begin{equation}
    \begin{split}
        \log p \left( \mathbf{x}_i \middle| \boldsymbol{\mu}^{(l)}, \boldsymbol{\Sigma}^{(l)} \right) = -\frac{d}{2}\log 2 \pi -  \frac{1}{2} \log \left| \boldsymbol{\Sigma}^{(l)} \right| - \frac{1}{2} \left( \mathbf{x}_i - \boldsymbol{\mu}^{(l)} \right)^\top \boldsymbol{\Sigma}^{(l)^{-1}} \left( \mathbf{x}_i - \boldsymbol{\mu}^{(l)} \right).
    \end{split}
\end{equation}
In the maximization stage, the mean and variance of each cluster that maximize the log-likelihood have an analytical solution that is,
\begin{equation}
    \begin{split}
        \boldsymbol{\mu}^{(l)} &= \frac{\sum_{i = 1}^N \gamma_i^{(l)} \mathbf{x}_i}{\gamma^{(l)}} \\ 
        \boldsymbol{\Sigma}^{(l)} &= \frac{\sum_{i = 1}^N \gamma_i^{(l)} \mathbf{x}_i \mathbf{x}_i^\top}{\gamma^{(l)}} - \boldsymbol{\mu}^{(l)} \boldsymbol{\mu}^{(l)^\top}.
    \end{split}
\end{equation}

The temperatures or heights can be approximated with univariate normal distribution. The extension of the GMM is the same for the case of one variable or multiple variables. The theory behind mixture models, as well as the EM algorithm, is fully developed in \cite{MURPHY2012}. 

\subsection{Bayesian Metrics}\label{subsec:BIC}

BIC is a metric to choose between models but penalizing the models that have higher number of parameters, and have more samples \cite{SCHWARZ1978}. The BIC in a mixture model is,
\begin{equation}
    \begin{split}
        \mathcal{BIC} ( \hat{\boldsymbol{\theta}} )_L &= \lambda \log N - 2 \log \mathcal{Q} ( \hat{\boldsymbol{\theta}} ) \\
        &= \lambda \log N - 2 \left\{ \sum_{l = 1}^L \sum_{i = 1}^N  \mathbb{I} \left(z_i = l\right) \left[ \log  \pi^{(l)} + \log p \left( x_i \middle| \hat{\boldsymbol{\theta}}^{(l)} \right) \right] \right\},
    \end{split}
\end{equation}
where $\lambda$ is the number of parameters in the model, and $N$ is the number of samples. As a pixel is assumed to be in one wind layer or another $\mathbb{I} (z^{(l)}_i = l) \in \{0, 1\}$.

The BIC is close related to Akaike Information Criterion (AIC) \cite{AKAIKE1974},
\begin{equation}
    \mathcal{AIC} ( \hat{\boldsymbol{\theta}} )_L = 2 \lambda - 2 \log \mathcal{Q} ( \hat{\boldsymbol{\theta}} ),
\end{equation}
In other metrics, such as the Classification Likelihood Criterion (CLC),
\begin{equation}
    \mathcal{CLC} ( \hat{\boldsymbol{\theta}} )_L = 2 \mathcal{H} ( \hat{\boldsymbol{\theta}} ) - 2 \log \mathcal{Q} ( \hat{\boldsymbol{\theta}} ),
\end{equation}
uses the entropy function $\mathcal{H} ( \cdot )$ in the context of information theory. CLC is similar to the AIC \cite{NGUYEN2016}, but applying the entropy as a penalizing factor instead of the number of parameters. The entropy in a mixture model is,
\begin{equation}
    \mathcal{H} ( \hat{\boldsymbol{\theta}} )_L = \sum^L_{l = 1} \sum^N_{k = 1} \gamma_i^{(l)} \log \gamma_i^{(l)}.
\end{equation}
The ICL, which is
\begin{equation}
    \mathcal{ICL} ( \hat{\boldsymbol{\theta}} )_L = \mathcal{BIC} ( \hat{\boldsymbol{\theta}} )_L + 2 \mathcal{H} ( \hat{\boldsymbol{\theta}} ),
\end{equation}
is based on both BIC and the entropy.

The number of clusters $L$ and the likelihood function, is different in each model $\mathcal{M}_L$, thus each model is expected to have a different $\mathcal{BIC} ( \hat{\boldsymbol{\theta}} )_L$, $\mathcal{AIC} ( \hat{\boldsymbol{\theta}} )_L$, $\mathcal{CLC} ( \hat{\boldsymbol{\theta}} )_L$, and $\mathcal{ICL} ( \hat{\boldsymbol{\theta}} )_L$. For all those metrics, the optimal number of clusters is when the value of the metric is the lowest.

\subsection{Hidden Markov Model}

A HMM is state space model which latent variables (i.e. system states) are discrete. In the problem of detecting the number of wind velocity fields in an image, a HMM is implemented to infer the cluster number $L$ in a mixture model. We assume that the current state of the system $L_t$ (i.e. cluster number) is a Markov process conditional to the previous observed states $p (L_t \mid L_{1}, \dots, L_{t - 1})$  \cite{BISHOP2006}. For simplification, we propose to model the process as a first-order Markov chain, which current state is only conditional to the immediately previous state $p (L_t \mid L_{1}, \dots, L_{t - 1}) = p (L_{t} \mid L_{t - 1}) $. Henceforth, $L$ is defined as the HMM hidden variable, which represents the number of detected wind velocity fields $L \in \{1, 2\}$ in image $L_t$, and $\mathbf{x}_{i,t}$ is an observations (i.e. feature vector).

The parameters $\boldsymbol{\theta}{}^{(l)}_t = \{ \pi{}^{(l)}_t, \boldsymbol{\mu}{}^{(l)}_t, \boldsymbol{\Sigma}{}^{(l)}_t \}$ of each distribution $l$ in the mixture model, and the hidden state of the system $L_t$ in image $t$ are the MAP estimation obtained applying the Bayes' theorem,
\begin{equation}
    \begin{split}
        \label{eq:bayes_theorem}
        p \left( \mathbf{z}_{i,t} \middle| \mathbf{x}_{i,t}, \boldsymbol{\Theta}_t \right) &= \frac{p \left( \mathbf{x}_{i,t}, \mathbf{z}_{i,t} \middle| \boldsymbol{\Theta}_t \right)}{p \left( \mathbf{x}_{i,t} \right)} \\
        &\propto p \left( \mathbf{x}_{i,t}, z^{(l)}_{i,t} \middle| \boldsymbol{\theta}_t^{(l)} \right) p \left( L_t \middle| L_{t - 1}  \right) \\
        &\propto p \left( \mathbf{x}_{i,t}, z^{(l)}_{i,t} \middle| \boldsymbol{\mu}_t^{(l)}, \boldsymbol{\Sigma}_t^{(l)} \right) p \left( \pi^{(l)}_t \middle| \boldsymbol{\alpha}^{(l)} \right) p \left( L_t \middle| L_{t - 1} \right).
    \end{split}
\end{equation}
The set with all the parameters in the mixture model in state $L_t$ are $\boldsymbol{\Theta}{}_t = \{ \pi{}^{(1)}_t, \boldsymbol{\mu}{}^{(1)}_t, \boldsymbol{\Sigma}{}^{(1)}_t, \dots, \pi{}^{(L_t)}_t, \boldsymbol{\mu}{}^{(L_t)}_t, \boldsymbol{\Sigma}{}^{(L_t)}_t  \}$, and the parameters of the prior distribution of the clusters weights $\pi^{(l)}$ are $\boldsymbol{\alpha}{}^{(l)}$.

The joint distribution $p ( \mathbf{x}_{i,t}, \mathbf{z}{}^{(l)}_{i,t} \mid \boldsymbol{\theta}{}^{(l)}_t ) = p ( \mathbf{x}_{i,t} \mid \mathbf{z}{}^{(l)}_{i,t}, \boldsymbol{\theta}^{(l)}_t ) p ( \mathbf{z}{}^{(l)}_{i,t} )$ is factorized to independently infer a mixture model of each feature,
\begin{equation}
    \begin{split}
        \label{eq:factorization}
        p \left( x_{i,t}^{(1)}, x_{i,t}^{(2)} \middle| z^{(l)}_{i,t}, \boldsymbol{\theta}_t^{(l)} \right) = p \left( x^{(1)}_{i,t} \middle| z^{(1,l)}_{i,t}, \boldsymbol{\theta}^{(l)}_{1,t} \right) p \left( x^{(2)}_{i,t} \middle|  x^{(1)}_{i,t}, z^{(1,l)}_{i,t}, z^{(2,l)}_{i,t}, \boldsymbol{\theta}^{(l)}_{2,t}\right) 
    \end{split}
\end{equation}
where $\boldsymbol{\theta}{}^{(l)}_t = \{ \mathbf{z}{}^{(l)}_{1,t}, \mathbf{z}{}^{(l)}_{1,t} \}$ and $\boldsymbol{\gamma}{}^{(l)}_t = \{ {z}{}^{(l)}_{1,t}, {z}{}^{(l)}_{2,t} \}$ are the parameters of the independent mixture models and the responsibilities respectively.

The posterior distribution of a mixture model is proportional to the joint probability of the observations $\mathbf{x}_{i,t}$ and the responsibilities $z{}^{(l)}_{i,t}$ of each cluster detected in image $t$. The state of the system $L_t$, which represents the number of clusters $L$ in the mixture models, is modelled using a HMM. Therefore, the mixture model parameters inferred using Eq. \eqref{eq:CDLL} is
\begin{equation}
    \label{eq:joint_distribution}
    p \left( \mathbf{X}_{t}, \mathbf{Z}^{(l)}_{t} \middle| \hat{\boldsymbol{\theta}}^{(l)}_t, \hat{\boldsymbol{\alpha}}^{(l)} \right) \triangleq \left\{ \sum^N_{i = 1} \left[ \prod^{L_t}_{l = 1} \left[  \pi^{(l)} p \left( \mathbf{x}_{i,t} \middle| \hat{\boldsymbol{\mu}}^{(l)}_t, \hat{\boldsymbol{\Sigma}}^{(l)}_t \right) \right]^{\mathbb{I} (z_{i,t} = l)} \right] \right\} p \left(\pi^{(l)} \middle| \hat{\boldsymbol{\alpha}}^{(l)}\right),
\end{equation}
where $\hat{\boldsymbol{\theta}}{}^{(l)}_t$ and $\hat{\boldsymbol{\alpha}}{}^{(l)}$ are the parameters that maximize the CDLL of the mixture model with $L_t$ clusters. 

The prior on the latent variable $L_t$, is defined as a distribution of the exponential family, 
\begin{equation}
    \label{eq:prior}
    \log p \left( L_t \middle| L_{t - 1} \right) = \frac{1}{Z} \exp \left[ - \psi \left( L_t, L_{t - 1}\right) \right],
\end{equation}
where the exponent $\psi ( L_t, L_{t - 1})$, is a function that depends on the previous state of the system $L_{t - 1}$,
\begin{equation}
    \psi \left( L_t, L_{t - 1} \right) \triangleq 
    \left\{
    \begin{array}{lll}
        -\beta & \text{if} & L_t = L_{t-1} \\ 
        +\beta & \text{if} & L_t \neq L_{t-1}
    \end{array} \right.
    \label{eq:hmm_prior},
\end{equation}
the parameter $\beta$ has to be cross-validated.

Combining Eq. \eqref{eq:joint_distribution} and Eq. \eqref{eq:prior} in Eq. \eqref{eq:bayes_theorem}, and taking logarithms and expectations with respect to $z_{i,t}$, the CDLL of a image $t$ being in state $L^{(t)}$ are,
\begin{equation} 
    \begin{split}
        \mathcal{Q} \left( \boldsymbol{\theta}, \boldsymbol{\theta}_{t - 1} \right) &= \left[ \sum^N_{i = 1} \sum^{L_t}_{l = 1} \gamma^{(l)}_{i,t} \log \pi^{(l)}_{t} + \sum^N_{i = 1} \sum^{L_t}_{l = 1} \gamma^{(l)}_{i,t} \log p \left( \mathbf{x}_{i,t} \middle| \boldsymbol{\theta}^{(l)}_t\right) \right] \\
        &\quad \quad \quad + \log p \left(\pi^{(l)} \middle| \hat{\boldsymbol{\alpha}}^{(l)}\right) - \psi \left( L_t, L_{t - 1} \right) + \mathrm{constant}.
    \end{split}
\end{equation}
It should be noted that $\psi \left( L_t, L_{t - 1} \right)$ is a constant with respect to $z_{i, t}$, so maximizing this equation is equivalent to maximizing the original CDLL in Eq. \eqref{eq:joint_distribution}.

After completing the inference of the mixture model parameters $\hat{\boldsymbol{\theta}}{}^{(l)}_t$ and $\hat{\boldsymbol{\alpha}}{}^{(l)}$ when $L_t = 1$ and $L_t = 2$, the optimal state of the system $\hat{L}_t \in \{ 1, 2\}$ is the MAP estimation obtained from,
\begin{equation}
    \hat{L}_t = \underset{L_t \in \{1, 2 \}}{\operatorname{argmax}} \ \left[ \sum_{i = 1}^N \sum_{l = 1}^{L_t} \log p \left( \mathbf{z}_{i,t} \middle| \mathbf{x}_{i,t}, \hat{\boldsymbol{\theta}}{}^{(l)}_t, \hat{\boldsymbol{\alpha}}{}^{(l)}, L_t \right) \right] - \psi \left( L_t, L_{t - 1} \right),
\end{equation}
where the latent variable $L_t$ defines the number of different wind velocity fields detected in an image.

\section{Experiments}

\subsection{Study Area and Data Acquisition System}

The climate of Albuquerque, NM is arid semi-continental with little precipitation, which is more likely during the summer months. The average altitude of the city is $1,620$m. Between mid May and mid June, the sky is clear or partly cloudy 80\% of the time. Approximately 170 days of the year are sunny, with less than 30\% cloud coverage, and 110 are partly sunny, with 40\% to 80\% cloud coverage. Temperatures range from a minimum of $268.71$K in winter to a maximum of $306.48$K in summer. Combined rainfall and snowfall are approximately $27.94$cm per year.

The proposed detection methods utilize data acquired by a DAQ system equipped with a solar tracker that updates its pan and tilt every second, maintaining the Sun in a central position in the images throughout the day. The IR sensor is a Lepton\footnote{https://www.flir.com/} radiometric camera with wavelength from 8 to 14 $\mu$m. The pixels in a frame are temperature measurements in centikelvin. The resolution of the IR images is $80 \times 60$ pixels. The DAQ is located on the roof area of the UNM-ECE building in Albuquerque, NM. The dataset composed of GSI measurements and IR images is available in a repository \cite{TERREN2020d}.

The weather features that were used to compute cloud height as well as to remove cyclostationary artifacts from the IR images are: atmospheric pressure, air temperature, dew point and humidity. The weather station measures every 10 minutes. The data was interpolated to match the IR images samples. The weather station is located at the University of New Mexico Hospital. It is publicly accessible\footnote{https://www.wunderground.com/dashboard/pws/KNMALBUQ473}.

\subsection{Image Preprocessing}

\begin{figure}[!htb]
    \centering
    \includegraphics[scale = 0.325]{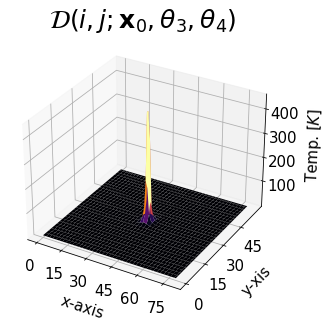}
    \includegraphics[scale = 0.325]{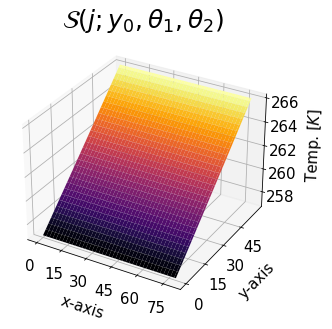}
    \includegraphics[scale = 0.325]{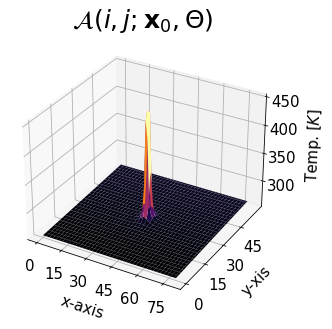}
    \includegraphics[scale = 0.325]{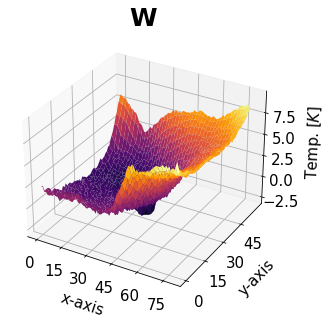}
    \caption{Irradiance effect models applied in the image preprocessing. From left to right, direct irradiance effect of the Sun, scattering effect produced by the atmosphere, and combination of the irradiance effects of the Sun and the atmosphere, and scattering effect produced by the germanium camera window.}
    \label{fig:atmospheric_models}
\end{figure}

\begin{figure}[!htb]
    \begin{minipage}{\linewidth}
        \includegraphics[scale = 0.45]{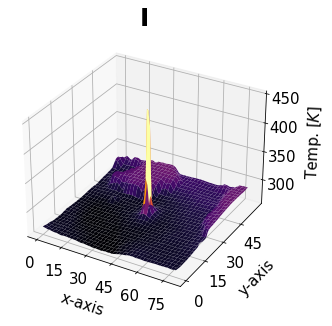}
        \includegraphics[scale = 0.45]{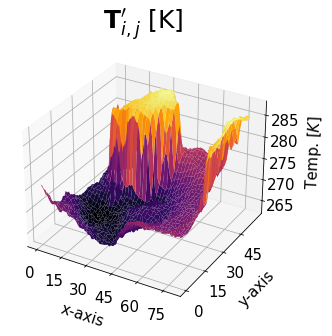}
        \includegraphics[scale = 0.45]{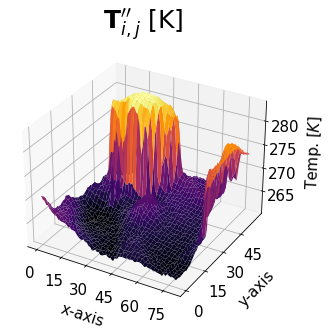}
    \end{minipage}
    \begin{subfigure}{\linewidth}
        \centering
        \includegraphics[scale = 0.325]{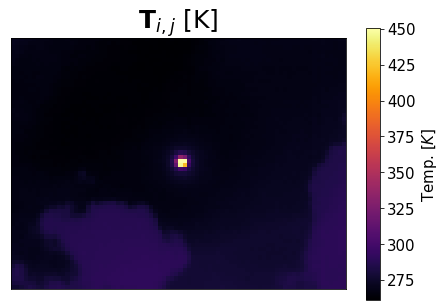}
        \includegraphics[scale = 0.325]{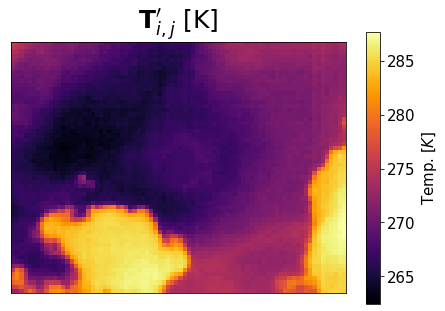}
        \includegraphics[scale = 0.325]{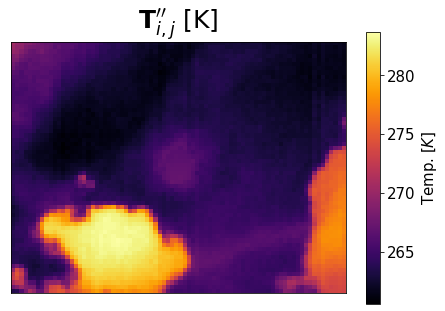}
    \end{subfigure}
    \caption{Preprocessing of the IR images. The images in the first row show the 3-dimensional image of the images in the second row. From left to right, raw IR image, IR after removing the effects of the atmospheric irradiance, and IR after removing the effects of the atmospheric irradiance and the scattering effect produced by the camera window. The models applied are shown in Fig. \ref{fig:atmospheric_models}.}
    \label{fig:preprocessing}
\end{figure}

The IR images were preprocessed to remove the effects of the direct irradiance from the Sun, the scattered irradiance from the atmosphere and the scattered irradiance from the germanium IR camera window (see Fig. \ref{fig:atmospheric_models}). The effect of the direct irradiance from the Sun is constant on the IR images, and is modelled and removed. The scattering effect produced by the atmosphere is cyclostationary, so the optimal model in each frame is different. The parameters of the atmospheric irradiance model depend on the azimuth and elevation angles of the Sun, and on weather features. The scattering effect produced by the germanium IR camera window is modelled and removed using the median IR image of the last in a set of clear-sky images (see Fig. \ref{fig:preprocessing}). The image processing methods and atmospheric conditions model are fully described in \cite{TERREN2021a}.

\begin{figure}[!htb]
    \centering
    \includegraphics[scale = 0.35]{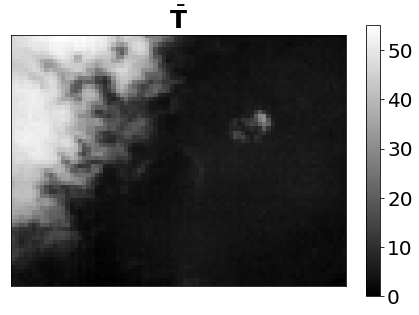}
    \includegraphics[scale = 0.35]{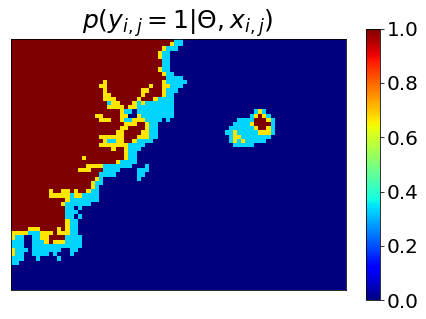}
    \includegraphics[scale = 0.315]{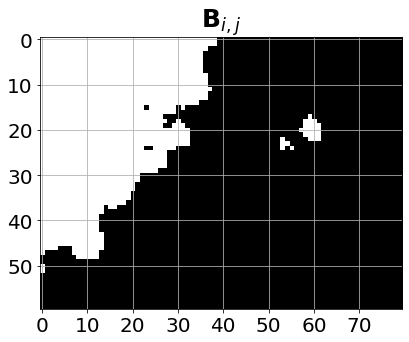}
    \caption{Cloud segmentation in IR images. The first image shows an IR image normalized to 8bits, the second image shows the probabilities computed by the segmentation algorithm of pixels belonging to a cloud, and the third image shows the segmentation after applying a $\geq 0.5$ threshold to the probabilities. The normalized images are also used to compute the velocity vectors in the proposed WLK.}
    \label{fig:segmentation}
\end{figure}

The proposed algorithm for the detection of clouds in multiple wind velocity fields requires that pixels containing clouds be previously segmented in the images. In this way, only features from clouds containing pixels are analyzed. The cloud segmentation algorithm implemented in this investigation is a voting scheme that uses three different cloud segmentation models. The segmentation models are a Gaussian process, a support vector machine and unsupervised Markov Random Field (see Fig. \ref{fig:segmentation}). The cloud segmentation models and feature extraction are explained in \cite{TERREN2020c}.

\subsection{WLK Parameters Cross-Validation}

A series of images with clouds flowing through different directions were simulated to cross-validate the set of parameters for each one of the mentioned methods \cite{TERREN2020a}. The WLK method was found to be the most suitable for this application. The investigation was searching for a dense implementation of a motion vector method to approximate the dynamics of a cloud. The most suitable method was found to be WLK. The optimal window size, WLS regularization, and temporal kernel amplitude are: $W = 16 \left[ \mathrm{pixels}^2\right]$, $\tau = 1 \times 10^{-8}$ and $\sigma = 1$.

\subsection{Mixture Models Parameters Cross-Validation}

The parameters that have to be cross-validated for each mixture model are $\alpha_l$ and $\beta$. The parameter $\alpha_l$ in Eq. \eqref{eq:mm_prior} is the parameter of the prior distribution of the cluster weight $\pi_l$ in a mixture model. The parameters $\boldsymbol{\alpha}$ in a mixture model of the temperatures are cross-validated as $\boldsymbol{\alpha} \triangleq \alpha_0 \mathbf{1}_{L \times 1}$ for simplification. Equivalently, the parameters $\boldsymbol{\alpha}$ in a mixture model of the velocity vectors are cross-validated as $\boldsymbol{\alpha} \triangleq \alpha_1 \mathbf{1}_{L \times 1}$. The parameter $\beta$ in Eq. \eqref{eq:hmm_prior} is the prior of the number of cloud layers in an image used in the sequential HMM. Both in training and in testing, the state $L_t$ is initialized to the opposite number of cloud layers in the IR image sequence (e.g. if $L_t = 2$ in the sequence, $L$ is initialized as $L_0 = 1$).

The parameter's cross-validation was implemented using a High Performance Computer (HPC). Even when the parameter's cross-validation is implemented in a HPC, it is still computationally expensive and the number of validation samples is prohibitive. The  cross-validation used two nodes and it was distributed across sixteen cores, which corresponds with the number of possible $\beta$ in the cross-validation $\beta = \{ 0, \dots, 1000\}$. In each core, all possible combinations of $\alpha_0$, which is the parameter of the priors of the clusters corresponding to temperature, and $\alpha_1$ (velocity vector cluster priors) were cross-validated $\alpha_l = \{ 1, \dots, 1000\}$. The performance of each combination of parameters for each mixture model were evaluated in the six training sequences. The optimal combination of parameters for each mixture model is the one which achieved the highest accuracy.


The training dataset is formed by six sequences of 21 consecutive images acquired on six different days. The training sequences were captured on different seasons and different times of the day. IR images were manually labelled as having one cloud layer $L = 1$ or two cloud layers $L = 2$. The images from three of the six days show a layer of cirrostratus in winter during the morning, altostratus in spring during the afternoon and stratocumulus in the summer during the afternoon. The other three days show two layers of altostratus and stratocumulus in winter at noon, cirrostratus and altocumulus in spring during the afternoon, and cirrocumulus and cumulus in summer during the morning. As the proposed method is an online machine learning algorithm, the training dataset is used only for validating the prior distribution parameters. The optimal parameters of the mixture models are computed for each new sample during the implementation.

\subsection{Testing Performance}

The testing dataset is composed of ten consecutive sequences of 30 images. The sequences were acquired at different hours of the day and in different seasons. The images in the testing dataset were acquired after the training dataset. The images in the testing dataset were manually labelled in the same way as the training set. The testing dataset includes five sequences of images that have one layer of clouds, and five sequences of images that have two layers of clouds. The clouds in the sequences with one layer are: stratocumulus on a summer morning, cumulus on a summer morning, stratus on a summer afternoon, cumulus on a fall morning, and stratocumulus on a winter morning. The clouds in the sequences with two layers are: cumulus and cirrostratus on a summer morning, cumulus and altostratus on a fall morning, cumulus and cirrus on a fall afternoon, stratucumulus and altostratus on a fall morning, and cumulus and nimbostratus on a winter afternoon.

\begin{table}
    \centering
    \scriptsize
    \setlength{\tabcolsep}{3.75pt} 
    \renewcommand{\arraystretch}{1.25} 
    \begin{tabular}{lcccccccccccccc}
        \toprule
        \multicolumn{10}{c}{Multiple Cloud Layers Detection Accuracy [\%]} \\
        \midrule
        \multicolumn{1}{c}{Univariate} & \multicolumn{5}{c}{Bayesian Metrics} & \multicolumn{3}{c}{Cross-Validated} & MAP-MRF \\
        \multicolumn{1}{c}{Likelihoods} & $\mathcal{Q}$ &  $\mathcal{BIC}$ & $\mathcal{AIC}$ & $\mathcal{CLC}$ & $\mathcal{ICL}$ & $\alpha_0$ & $\alpha_1$ & $\beta$ & $\psi \left( y^{(t)} = y^{(t - 1)} \right)$ \\
        \midrule
        $\mathcal{B} ( \bar{T}_{i,j} )$ & 66.67 & 70.74 & 70.74   & 76.3 &  76.67 & 1 & - & 200 & 75.93 \\
        $\mathcal{G} ( \tilde{T}_{i,j} ) $ & 60 & 61.11 & 60.74 & 66.67 & 70.74 & 1 & - & 300 & 53.55 \\
        $\mathcal{N} ( T_{i,j} ) $ & $\mathbf{74.07}$ & $\mathbf{73.7}$ & $\mathbf{71.85}$ & $\mathbf{79.63}$ & $\mathbf{81.11}$ & 100 & - & 500 & $\mathbf{87.41}$ \\
        $\mathcal{VM} ( \phi_{i,j} ) $ & 57.41 & 60.37 & 61.85 & 58.52 & 61.11 & - & 1 & 500 & 60 \\
        $\mathcal{G} ( r_{i,j} ) $ & 41.48 & 40.37 & 42.22 & 39.63 & 36.67 & - & 10 & 1000 & 37.78 \\
        \bottomrule
    \end{tabular}
    \caption{Detection accuracy achieved when univariate probability functions are used in a mixture model likelihood. The detection accuracy is compared using different Beyesian metrics and a mixture model with a prior on the weights and the clusters number.}
    \label{tab:wlk_univariate_mms}
\end{table}

\begin{table}
    \centering
    \scriptsize
    \setlength{\tabcolsep}{1.35pt} 
    \renewcommand{\arraystretch}{1.25} 
    \begin{tabular}{lcccccccccccccc}
        \toprule
        \multicolumn{10}{c}{Multiple Cloud Layers Detection Accuracy [\%]} \\
        \midrule
        \multicolumn{1}{c}{Multivariate} & \multicolumn{5}{c}{Bayesian Metrics} & \multicolumn{3}{c}{Cross-Validated} & MAP-MRF \\
        \multicolumn{1}{c}{Likelihoods} & $\mathcal{Q}$ &  $\mathcal{BIC}$ & $\mathcal{AIC}$ & $\mathcal{CLC}$ & $\mathcal{ICL}$ & $\alpha_0$ & $\alpha_1$ & $\beta$ & $\psi \left( y^{(t)} = y^{(t - 1)} \right)$ \\
        \midrule
        $\mathcal{N} ( \mathbf{v}_{i,j} ) $ & 65.93 & 68.15 & 67.04 & 69.63 & 67.41 & - & 100 & 600 & 61.85 \\
        $\mathcal{N} ( T_{i,j}, \mathbf{v}_{i,j} )$ & 49.63 & 52.96 & 51.11 & 55.93 & 62.96 & 1000 & - & 1000 & 55.19 \\
        $\mathcal{BG} ( \tilde{T}_{i,j}, r_{i,j} )$ & 50 & 50.37 & 50 & 50.37 & 50.74 & 1 & - & 0 & 50 \\
        $\mathcal{BG} ( \tilde{T}_{i,j}, r_{i,j} ) \mathcal{VM} ( \phi_{i,j} )$ & 50 & 50 & 50 & 48.89 & 51.11 & 1 & 1 & 0 & 50 \\
        $\mathcal{B} ( \bar{T}_{i,j} ) \mathcal{N} ( \mathbf{v}_{i,j} ) $ & $\mathbf{70}$ & $\mathbf{71.11}$ & $\mathbf{70.37}$ & $\mathbf{75.56}$ & $\mathbf{75.93}$ & 10 & 1000 & 650 & $\mathbf{94.44}$ \\
        $\mathcal{G} ( \tilde{T}_{i,j} ) \mathcal{N} ( \mathbf{v}_{i,j} )$ & 61.48 & 61.48 & 62.96 &  55.19 & 57.04 & 1 & 100 & 450 & 68.15 \\
        \bottomrule
    \end{tabular}
    \caption{Detection accuracy achieved when multivariate probability functions are used in the likelihood of the mixture model. The detection accuracy is compared using different Bayesian metrics and a maximum a posteriori implementation of a mixture model with a prior on the weights and cluster numbers.}
    \label{tab:wlk_multivariate_mms}
\end{table}

\begin{table}
    \centering
    \scriptsize
    \setlength{\tabcolsep}{1.35pt} 
    \renewcommand{\arraystretch}{1.25} 
    \begin{tabular}{lcccccccccccccc}
        \toprule
        \multicolumn{10}{c}{Multiple Cloud Layers Detection Accuracy [\%]} \\
        \midrule
        \multicolumn{1}{c}{Factorized} & \multicolumn{5}{c}{Bayesian Metrics} & \multicolumn{3}{c}{Cross-Validated} & MAP-MRF \\ \multicolumn{1}{c}{Univariate Likelihoods} & $\mathcal{Q}$ &  $\mathcal{BIC}$ & $\mathcal{AIC}$ & $\mathcal{CLC}$ & $\mathcal{ICL}$ & $\alpha_0$ & $\alpha_1$ & $\beta$ & $\psi \left( y^{(t)} = y^{(t - 1)} \right)$ \\
        \midrule
        $\mathcal{B} ( \bar{T}_{i,j} ) \mathcal{VM} ( \phi_{i,j} ) $ & 68.89 & 68.89 & 69.63 &  $\mathbf{73.33}$ & $\mathbf{74.44}$ & 1 & 10 & 650 & $\mathbf{97.41}$ \\
        $\mathcal{B} ( \bar{T}_{i,j} ) \mathcal{G} ( r_{i,j} ) $ & 51.48 & 48.89 & 49.26 & 52.96 & 49.63 & 1 & 1000 & 650 & 69.26 \\
        $\mathcal{B} ( \bar{T}_{i,j} ) \mathcal{VM} ( \phi_{i,j} ) \mathcal{G} ( r_{i,j} ) $ & 55.19 & 55.19 &  55.56 & 57.41 & 56.3 & 10 & 100 & 1000 & 88.15 \\
        $\mathcal{G} ( \tilde{T}_{i,j} ) \mathcal{VM} ( \phi_{i,j} ) $ & 50.37 &  50.37 &  62.96 & 58.89 & 60 & 10 & 1 & 400 & 67.78 \\
        $\mathcal{G} ( \tilde{T}_{i,j} ) \mathcal{G} ( r_{i,j} ) $ & 62.22 &  62.22 & 50  &  50.37  & 50.74  & 10 & 10 & 400 & 54.81 \\
        $\mathcal{G} ( \tilde{T}_{i,j} ) \mathcal{VM} ( \phi_{i,j} ) \mathcal{G}  ( r_{i,j} ) $ & 53.7 & 56.67 & 52.96 & 46.3 & 50.74 & 10 & 1 & 650 & 51.11 \\
        $\mathcal{N} ( T_{i,j} ) \mathcal{VM} ( \phi_{i,j} ) $ & $\mathbf{78.15}$ & $\mathbf{76.67}$ & $\mathbf{77.41}$ & 72.96  & 71.85 & 100 & 1000 & 600 & 85.93 \\  
        $\mathcal{N} ( T_{i,j} ) \mathcal{G} ( r_{i,j} ) $ & 64.07 & 66.67 & 64.44 & 61.85 &  70.37  & 1000 & 100 & 500 & 80.37 \\
        $\mathcal{N} (  T_{i,j}  ) \mathcal{VM} ( \phi_{i,j} ) \mathcal{G} ( r_{i,j} ) $ & 73.33 &  67.78 & 68.15 &  72.96 & 68.89 & 100 & 1 & 400 & 72.96 \\
        $\mathcal{VM} ( \phi_{i,j}) \mathcal{G} ( r_{i,j} ) $ & 51.11 & 48.52 & 48.15 & 42.22 & 43.70 & - & 10 & 1000 & 57.78 \\
        \bottomrule
    \end{tabular}
    \caption{Detection accuracy achieved when the mixture model likelihood is factorized in the product of independent likelihood functions for each feature. The detection accuracy is compared using different Bayesian metrics and adding a prior of the weights and the cloud layers number to the mixture model.}
    \label{tab:wlk_factorize_univariate_mm}
\end{table}

\begin{figure}[!htb]
    \begin{subfigure}{\linewidth}
        \ \ \ \ \ \ \includegraphics[scale = 0.26]{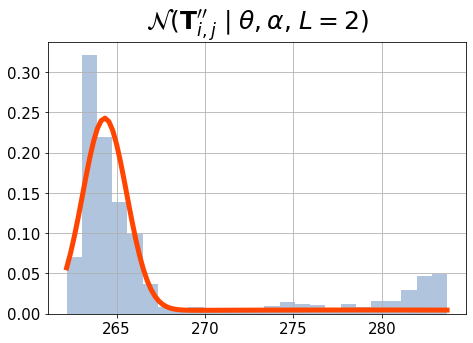}
        \ \ \ \ \ \ \includegraphics[scale = 0.26]{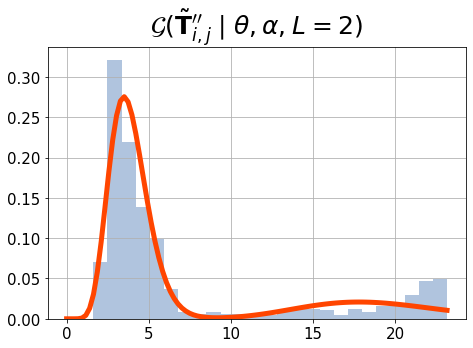}
        \ \ \ \ \ \ \ \ \ \includegraphics[scale = 0.26]{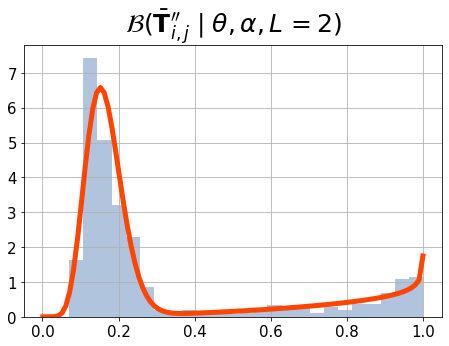}
    \end{subfigure}
    \begin{minipage}{\linewidth}
        \centering
        \includegraphics[scale = 0.325]{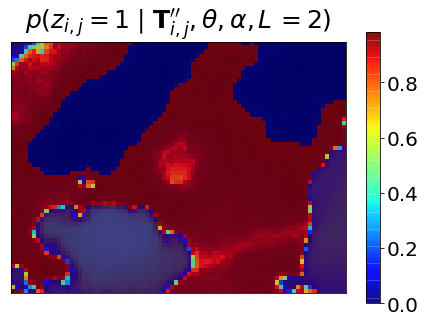}
        \includegraphics[scale = 0.325]{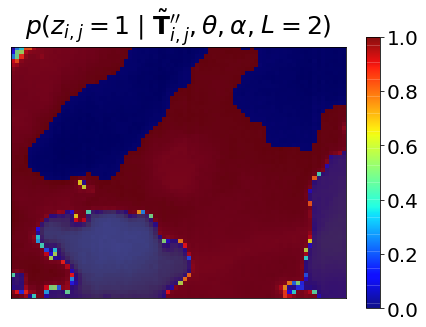}
        \includegraphics[scale = 0.325]{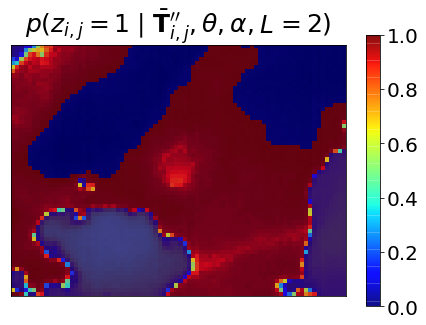}
    \end{minipage}
    \begin{minipage}{\linewidth}
        \centering
        \includegraphics[scale = 0.325]{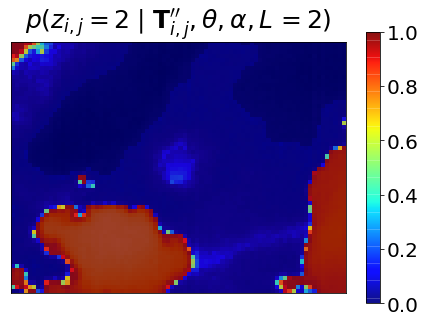}
        \includegraphics[scale = 0.325]{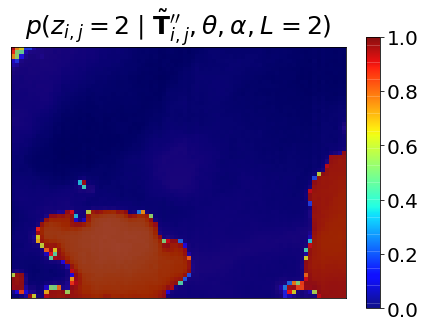}
        \includegraphics[scale = 0.325]{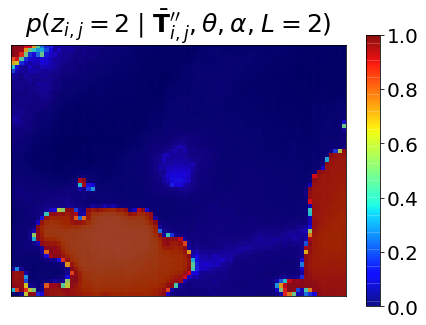}
    \end{minipage}
    \caption{First row shows the distribution of the temperatures inferred using a GMM, GaMM and BeMM. The mixture model likelihood is evaluated with the optimal parameters. The mixture model posterior probabilities for the top cloud layer and bottom cloud layer are shown in the middle and bottom rows respectively.}
    \label{fig:temp_dist}
\end{figure}

We assume that the distribution of the velocity vectors is different in each cloud layer that appears in an IR image. In addition, we assume that  a correlation exists between the height of a cloud and its velocity vectors. As the height of a cloud is a function of its temperature, we propose to use the temperature of the pixels and the velocity vectors to infer the number of cloud layers in an IR image. The distribution of temperatures is inferred using a BeMM, GaMM and GMM, see Fig. \ref{fig:temp_dist}. The performances of each distribution are analyzed and compared in tables \ref{tab:wlk_univariate_mms}-\ref{tab:wlk_multivariate_mms}. The posterior probabilities of the temperature mixture models in Fig. \ref{fig:temp_dist} are the weights used in the WLK.

\begin{figure}[!htb]    
    \centering
    \includegraphics[scale = 0.425]{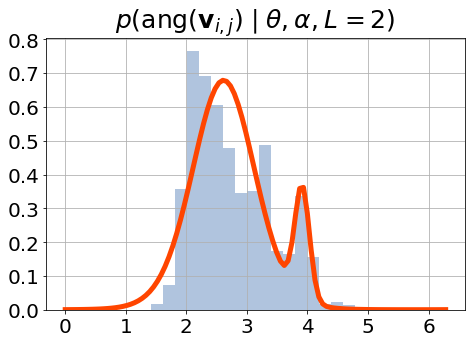}
    \includegraphics[scale = 0.425]{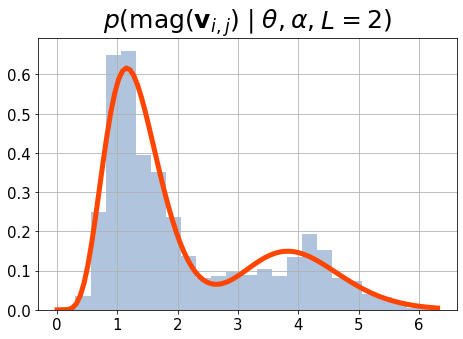}
    \caption{Distribution of the velocity vector's angle and magnitude. The mixture model likelihood is evaluated with the optimal parameters. The distribution of the velocity vector angle was inferred using a VMMM (left). The distribution of the velocity vector magnitude was inferred using GaMM (right).}
    \label{fig:vel_dist}
\end{figure}

The distribution of the velocity vector components is inferred using a multivariate GMM. The performance of the multivariate GMM is compared to the distribution of the velocity vectors magnitude and angle inferred factorizing the probability of the velocity vectors into two independent probability functions (see table \ref{tab:wlk_multivariate_mms}). In Fig. \ref{fig:vel_dist}, the distributions of the velocity angle and magnitudes are inferred using a VMMM  and GaMM respectively.

\begin{figure}[!htb]  
    \centering
    \includegraphics[scale = 0.425]{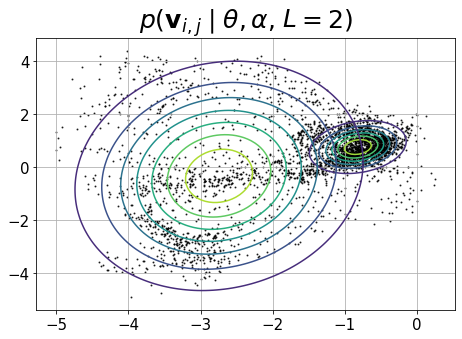}
    \includegraphics[scale = 0.425]{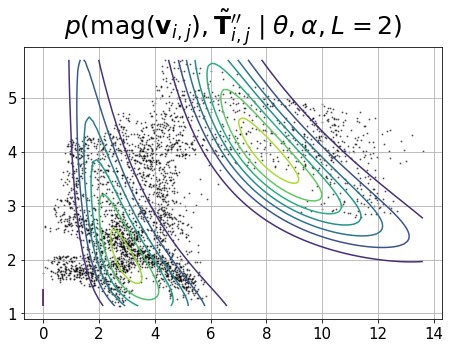}
    \includegraphics[scale = 0.475]{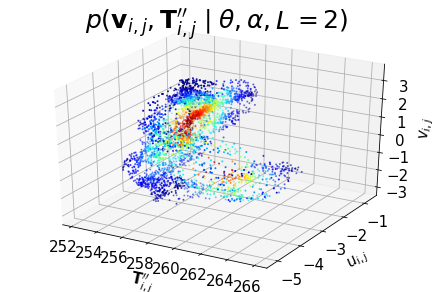}
    \caption{The distributions of the temperatures and the velocity vectors. The graphs show the density of mixture model likelihood evaluated with the optimal parameters. The distribution of the velocity vectors was inferred using a multivariate GMM (top left graph). The distribution of the temperatures and the velocity vector's magnitude was inferred using a BGaMM (top left graph). The distribution of the temperatures and the velocity vectors was inferred using a multivariate GMM (bottom graph).}
    \label{fig:temp_vel_dist}
\end{figure}

When weights are not applied to the LK method, the distribution of the temperatures, velocity vector angles and magnitude can be inferred using a multivariate GMM. Similarly, a DGaMM is also proposed to infer the distribution of the temperatures and velocity vector magnitudes. The multivariate GMM and DGaMM likelihood are displayed in Fig. \ref{fig:temp_vel_dist}. In this case, the probability of the velocity vector angles is factorized and inferred independently using a VMMM. The performance of these two mixture models are also shown in table \ref{tab:wlk_multivariate_mms}.

\begin{figure}[!htb]
    \begin{subfigure}{\linewidth}
        \centering
        \hspace{0.cm}
        \includegraphics[scale = 0.35]{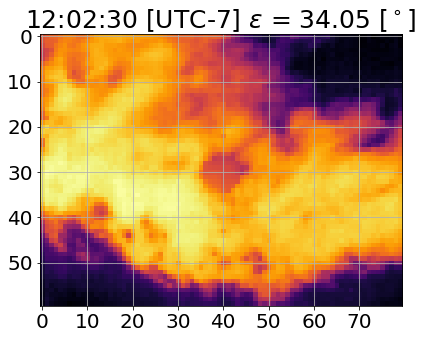}
        \hspace{0.cm}
        \includegraphics[scale = 0.35]{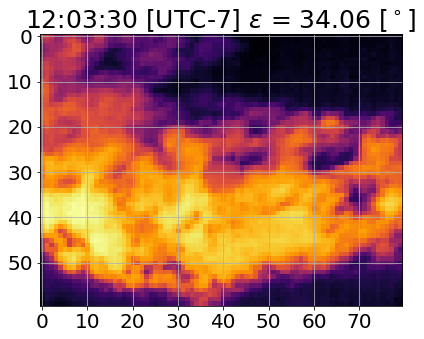}
        \hspace{0.05cm}
        \includegraphics[scale = 0.35]{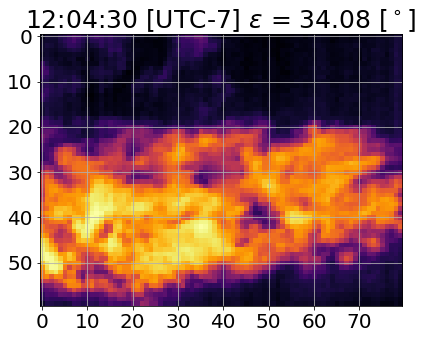}
    \end{subfigure}
    \begin{minipage}{\linewidth}
        \centering
        \hspace{0.cm}
        \includegraphics[scale = 0.35]{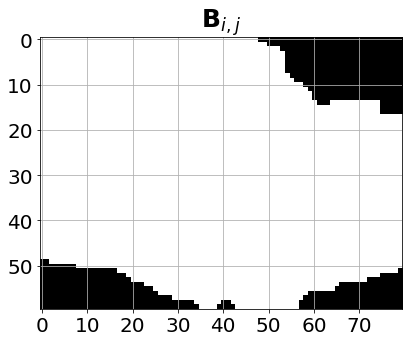}
        \hspace{0.15cm}
        \includegraphics[scale = 0.35]{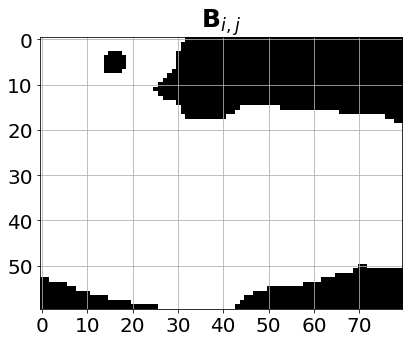}
        \hspace{0.2cm}
        \includegraphics[scale = 0.35]{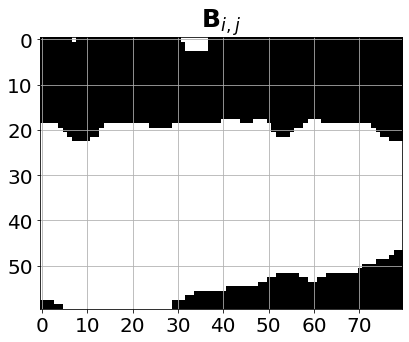}
    \end{minipage}
    \begin{minipage}{\linewidth}
        \centering
        \hspace{0.cm}
        \includegraphics[scale = 0.3]{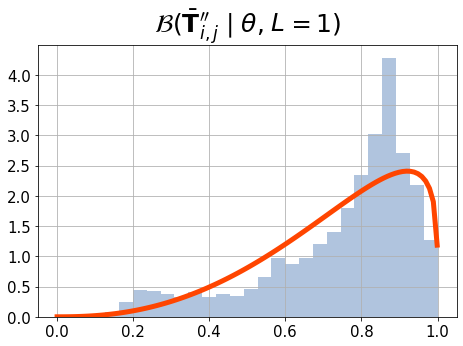}
        \hspace{0.3cm}
        \includegraphics[scale = 0.3]{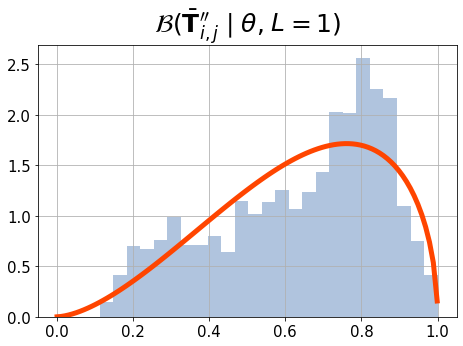}
        \hspace{0.425cm}
        \includegraphics[scale = 0.3]{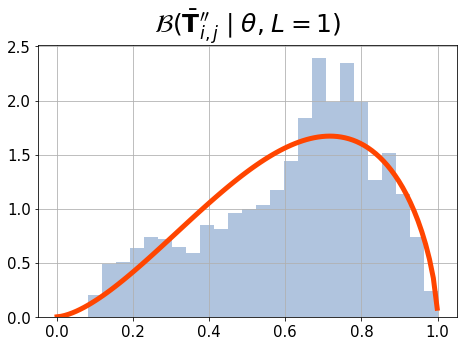}
    \end{minipage}
    \begin{minipage}{\linewidth}
        \centering
        \hspace{-0.025cm}
        \includegraphics[scale = 0.3]{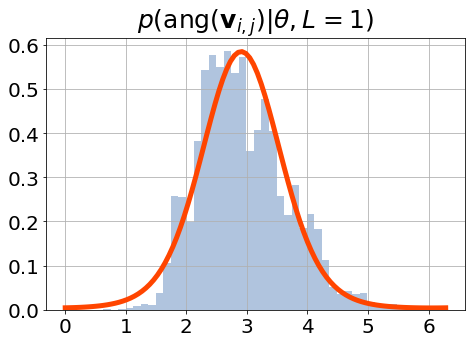}
        \hspace{0.25cm}
        \includegraphics[scale = 0.3]{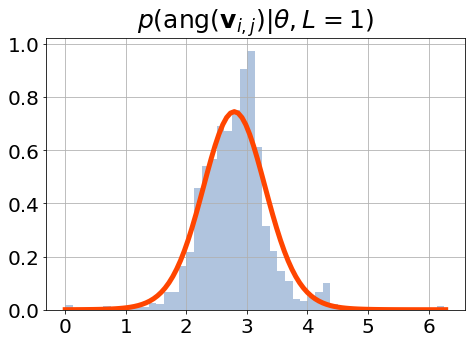}
        \hspace{0.35cm}
        \includegraphics[scale = 0.3]{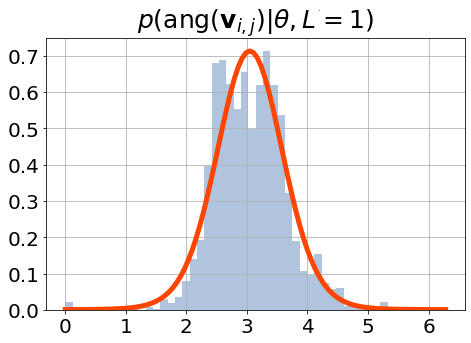}
    \end{minipage}
    \caption{Testing sequence of consecutive IR images acquired at 1 minute interval. In the first row, the IR images show a cloud flowing in a single detected wind velocity field. The images in the second row show the cloud segmentation. The graphs in the third and fourth row show the selected model distribution of the temperatures and the velocity vector angles respectively.}
    \label{fig:sample_1-2}
\end{figure}

\begin{figure}[!htb]
    \begin{subfigure}{\linewidth}
        \hspace{-0.1cm}
        \includegraphics[scale = 0.35]{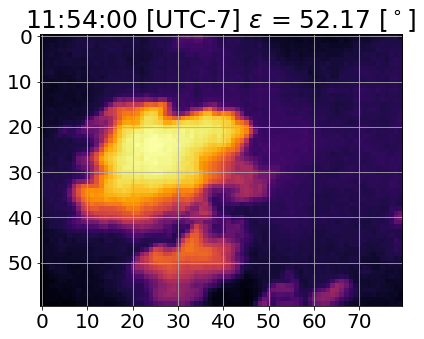}
        \hspace{0.075cm}
        \includegraphics[scale = 0.35]{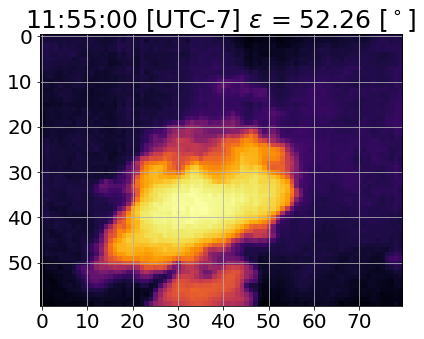}
        \hspace{0.2cm}
        \includegraphics[scale = 0.35]{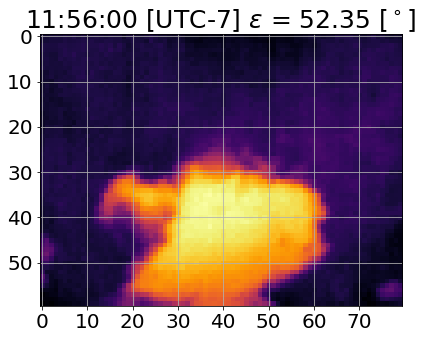}
    \end{subfigure}
    \begin{minipage}{\linewidth}
        \centering
        \includegraphics[scale = 0.35]{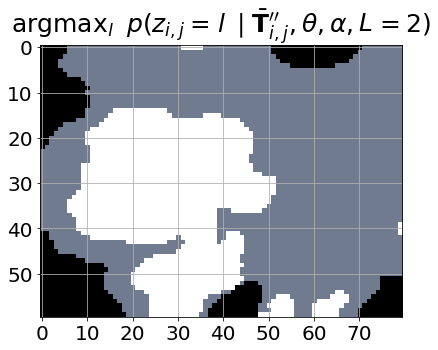}
        \includegraphics[scale = 0.35]{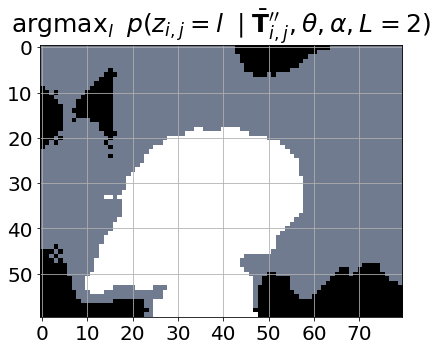}
        \hspace{0.cm}
        \includegraphics[scale = 0.35]{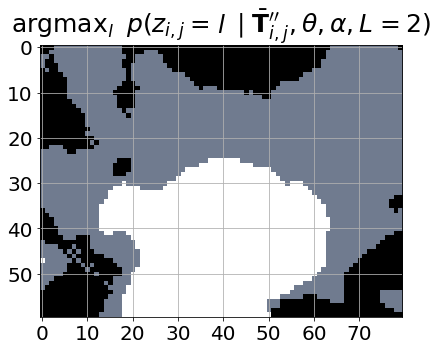}
    \end{minipage}
    \begin{minipage}{\linewidth}
        \hspace{0.25cm}
        \includegraphics[scale = 0.3]{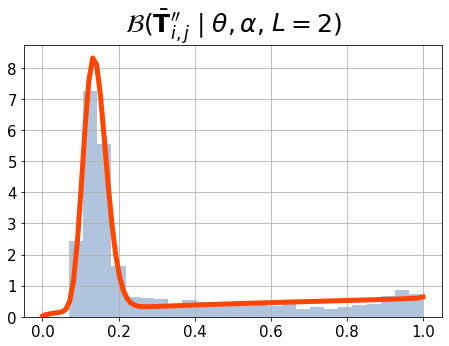}
        \hspace{0.5cm}
        \includegraphics[scale = 0.3]{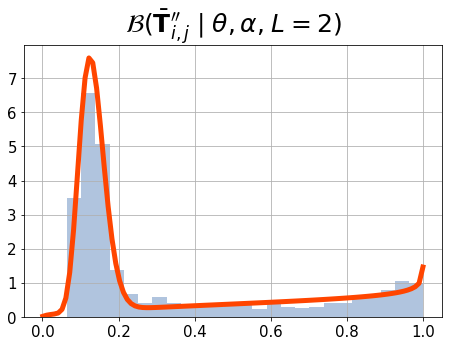}
        \hspace{0.65cm}
        \includegraphics[scale = 0.3]{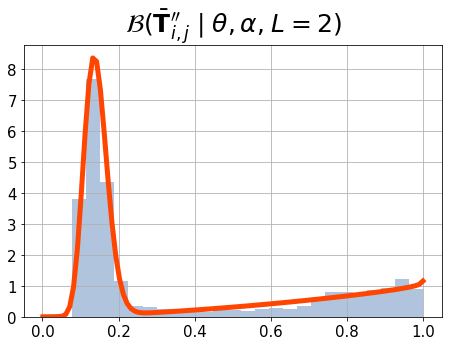}
    \end{minipage}
    \begin{minipage}{\linewidth}
        \hspace{0.025cm}
        \includegraphics[scale = 0.3]{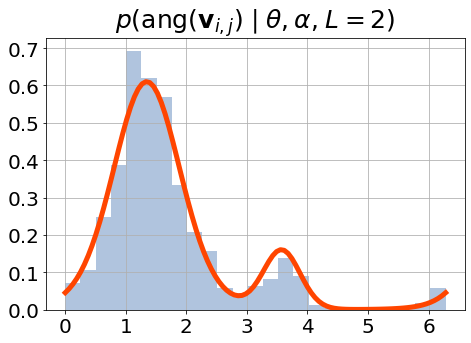}
        \hspace{0.25cm}
        \includegraphics[scale = 0.3]{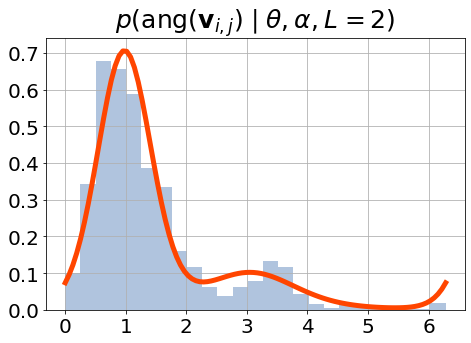}
        \hspace{0.4cm}
        \includegraphics[scale = 0.3]{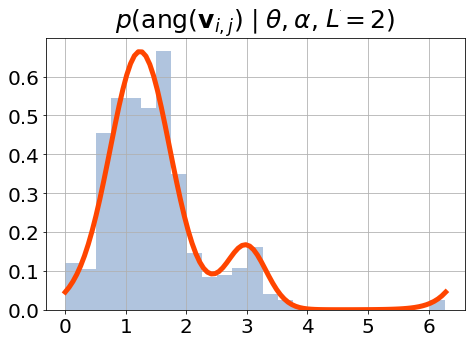}
    \end{minipage}
    \caption{Testing sequence of IR images with two detected cloud layers (first row). The time interval between images is 1 minute. The images in the second row show the pixels that belong to the low and high temperatures, in gray and white respectively. The pixels in black were classified as not belonging to a cloud by the segmentation algorithm. The third and fourth row show the distribution of the temperature and velocity vectors of the best model.}
    \label{fig:sample_2-1}
\end{figure}

\begin{figure}[!htb]
    \begin{subfigure}{\linewidth}
        \hspace{-0.1cm}
        \includegraphics[scale = 0.35]{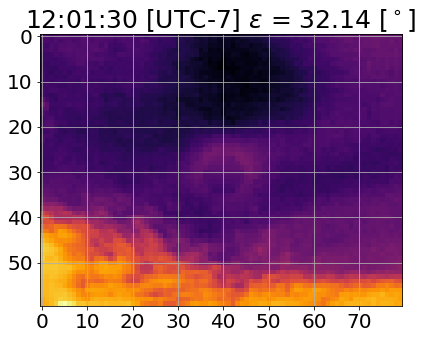}
        \hspace{0.1cm}
        \includegraphics[scale = 0.35]{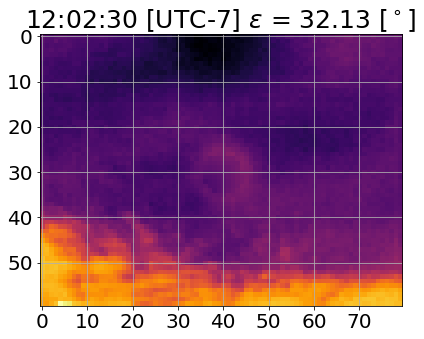}
        \hspace{0.2cm}
        \includegraphics[scale = 0.35]{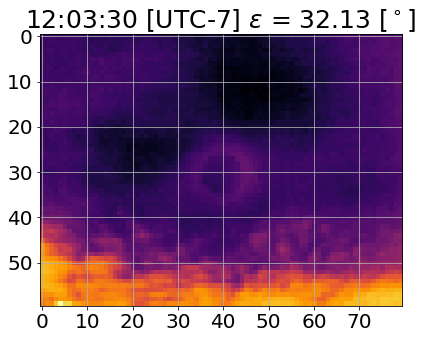}
    \end{subfigure}
    \begin{minipage}{\linewidth}
        \centering
        \includegraphics[scale = 0.35]{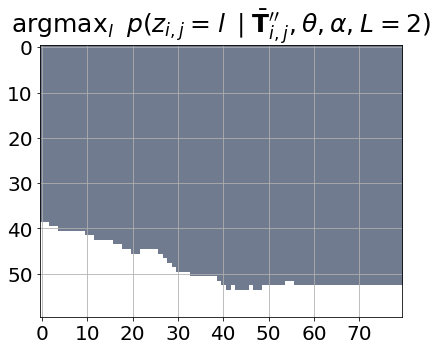}
        \includegraphics[scale = 0.35]{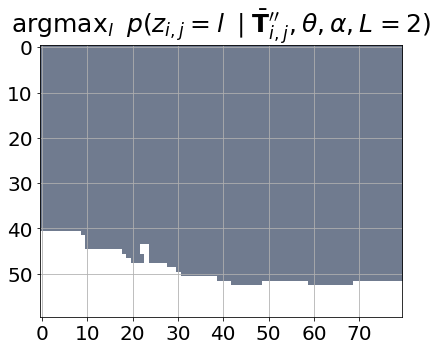}
        \hspace{0.cm}
        \includegraphics[scale = 0.35]{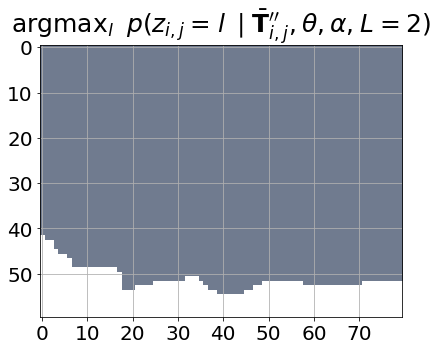}
    \end{minipage}
    \begin{minipage}{\linewidth}
        \hspace{0.2cm}
        \includegraphics[scale = 0.3]{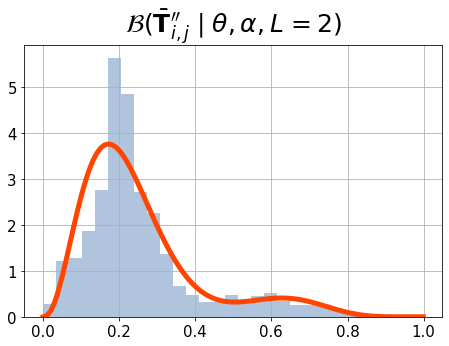}
        \hspace{0.6cm}
        \includegraphics[scale = 0.3]{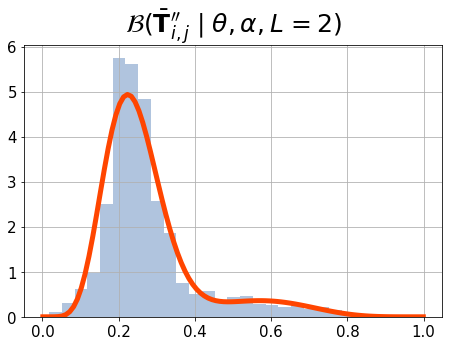}
        \hspace{0.6cm}
        \includegraphics[scale = 0.3]{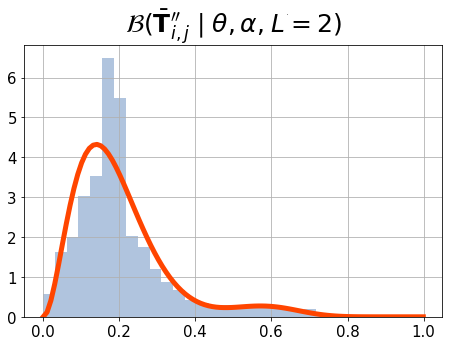}
    \end{minipage}
    \begin{minipage}{\linewidth}
        \hspace{-0.05cm}
        \includegraphics[scale = 0.3]{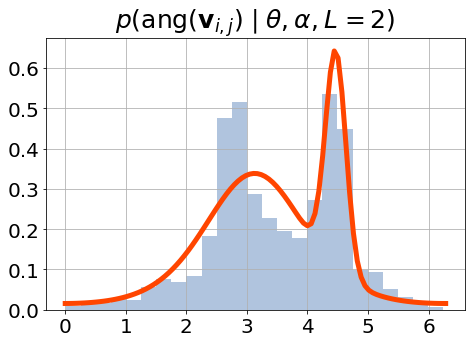}
        \hspace{0.375cm}
        \includegraphics[scale = 0.3]{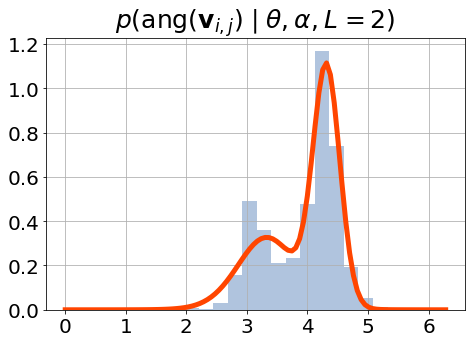}
        \hspace{0.375cm}
        \includegraphics[scale = 0.3]{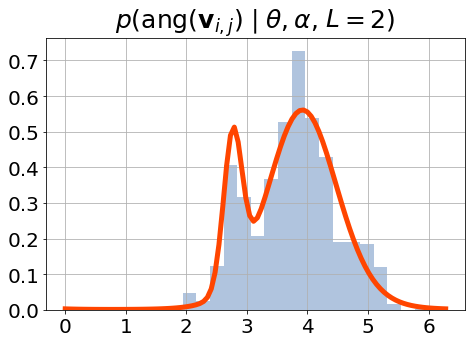}
    \end{minipage}
    \caption{Testing image with two detected cloud layers. The time interval between the IR image is 1 minute. The consecutive IR images are shown in the first row. The pixel labels of the top (grey) or the bottom cloud layer (white) are shown in the second row. The third and fourth row show the factorized likelihood that a VMMM and GaMM uses for the velocity vector angles and magnitude respectively.}
    \label{fig:sample_2-2}
\end{figure}

The experiments were carried out in the Wheeler high performance computer of UNM-CARC, which uses SGI AltixXE Xeon X5550 at 2.67GHz with 6 GB of RAM memory per core, has 8 cores per node, 304 nodes total, and runs at 25 theoretical peak FLOPS. It has Linux CentOS 7 installed.

\section{Discussion}


In the problem at hand, the BIC and AIC criteria do not produce an improvement on the detection accuracy with respect to accuracy achieved by ML criterion. The best detection accuracy achieved by a model that uses the ML criterion is $78.15 \%$. In contrast, the same model achieved a detection accuracy of $76.67 \%$ and $77.41 \%$ when the criteria were minimum BIC and AIC respectively (see table \ref{tab:wlk_factorize_univariate_mm}). This mixture model has a factorized likelihood which uses a normal probability function to infer the temperature distribution and a Von Mises to infer the velocity vector angles. However, when the minimum CLC and ICL criteria are applied, the detection accuracy improved with respect to that achieved by the ML criteria. The detection accuracy achieved by CLC, ICL and ML criteria were $81.11 \%$, $79.63 \%$ and $74.07 \%$ respectively (see table \ref{tab:wlk_univariate_mms}). Therefore, the best detection using a Bayesian metric was performed by a mixture model with a normal likelihood on the temperatures.

The detection accuracy of the proposed algorithm increases when the mixture model includes prior distributions on the mixture weights and the number of clusters. In these mixture models, the decision criterion is MAP. Adding a prior to the mixture weights and the cluster number is equivalent to regularizing the parameters. The prior adds certain known information to the model. In our problem, the prior on the number of  clusters produces the following effect:  if the previous frame had $L_{t - 1}$ clusters, the next frame is more likely to have $L_t$ as well. Similarly, the prior on the mixture weights assures that when the likelihood of two cloud layers is inferred, the cluster weights cannot vanish to zero. In table \ref{tab:wlk_univariate_mms}, when we look at the model that achieved the best detection accuracy using a Bayesian metric (ICL), the detection accuracy increased from $81.11 \%$ to $87.11 \%$. Nevertheless, the best detection accuracy using the MAP criterion reached $97.41 \%$ (see table \ref{tab:wlk_factorize_univariate_mm}). The model that  presents the best detection accuracy is a MAP mixture model with factorized likelihood which uses a beta probability function to infer the temperature distribution and a Von Mises distribution to infer the velocity vector angles. This validates our assumption that different cloud layers are at different heights (i.e. temperature) and hence the wind shear is also different (i.e. velocity vector angle). The proposed likelihood factorization allows us to find the optimal probability function of each feature independently.


The results show that it is feasible to identify different cloud layers in IR ground-based sky-images  (see Fig. \ref{fig:sample_1-2}-\ref{fig:sample_2-2}). The main advantage of this algorithm is that it provides the capability of independently estimating the motion of different cloud layers in an IR image using the posterior probabilities in Fig. \ref{fig:temp_dist}. This is useful in predicting when different clouds will occlude the Sun. The features and dynamics can be analyzed independently to increase the performances of a solar forecasting algorithm. Another advantage of the learning algorithm proposed is that it is unsupervised, so the user does not need to provide labels, which makes the training process automatic.


As it can be seen in tables \ref{tab:wlk_univariate_mms}-\ref{tab:wlk_factorize_univariate_mm}, the Bayesian metrics are not useful in this application. The highest accuracy achieved by a Bayesian metric was $81.11 \%$ with a GMM of the temperatures. The model selection was performed using minimum ICL criterion. The performances of the BGaMM are lower than the rest of the mixture models, thus it is not practical to assess the number of cloud layers in IR images. This is because the BGaMM tends to overfit even when the cluster weights are regularized using a prior distribution.

A disadvantage of the proposed unsupervised learning algorithm is that the EM requires several initializations to guarantee that the EM algorithm converges to the best local maxima. This is problematic when the cloud layer detection algorithm is meant for real-time applications. An implementation of the algorithm feasible in real-time applications will require multiple CPUs to run different initializations in parallel.

\section{Conclusions}


This investigation proposes an online unsupervised learning algorithm to detect moving clouds in different wind velocity fields. The mixture model of the pixel temperature is used to know when a cloud is below or on top of another. The posterior probabilities of the mixture model are used to compute velocity vectors. The algorithm to compute the velocity vectors is a weighted implementation of the Lucas-Kanade optical flow. The weights are the posterior probabilities of the mixture model. The velocity vectors are computed in a scenario that assumes one cloud layer and in another scenario that assumes  two cloud layers (using the cloud segmentation or the posterior probabilities respectively). The distribution of the velocity vectors and the temperatures is used to determine which one of the analyzed scenarios is the most likely. The proposed algorithm implements the MAP criterion.


The detection of clouds flowing in different wind velocity fields is useful to increase the accuracy of a forecasting algorithm that predicts the global solar irradiance that will reach a photovoltaic power plant. The prediction will aid a smart grid to adjust the generation mix to compensate for the decrease of energy generated by the photovoltaic panels.


In particular, the posterior probabilities of the pixel temperatures may aid the extraction of features using either image processing techniques, gradient-based learning (e.g. deep neural networks) or both. However, the posterior probabilities are only advantageous when there are multiple cloud layers in an IR image. The proposed method models a prior distribution of the cluster weights, and a prior function of each possible scenario. The prior function of the scenarios is a temporal implementation of a hidden Markov model. This investigation shows that the proposed method increases the detection accuracy compared to the accuracy achieved by the most common Bayesian metrics used in practice.


Future work in this area will implement cloud detection algorithms in a ramp-down and intra-hour solar forecasting algorithm. The dynamics of clouds may be analyzed independently to extract features from clouds moving in different wind velocity fields. The improvement in the performance can be assessed to determine how to combine the features extracted from different clouds to model their respective influence on the GSI that will reach the surface of a photovoltaic system. Another investigation may focus on the implementation of the proposed algorithm in images acquired using ground-based all-sky imagers that are sensitive to the visible light spectrum instead of the infrared. The most interesting aspect will be to fuse information acquired using visible and infrared light cameras.

\section{Acknowledgments}

This work has been supported by NSF EPSCoR grant number OIA-1757207 and the King Felipe VI endowed Chair. Authors would like to thank the UNM Center for Advanced Research Computing, supported in part by the National Science Foundation, for providing the high performance computing and large-scale storage resources used in this work.

\bibliographystyle{unsrt}  
\bibliography{mybibfile}

\end{document}